\documentclass[aps,notitlepage,oneside,superscriptaddress]{revtex4-1}
\usepackage{graphicx}
\usepackage{latexsym}
\usepackage{amssymb}
\usepackage{amsmath}
\usepackage{float}
\usepackage{pxfonts}
\usepackage{xcolor}
\definecolor{mygreen}{cmyk}{0.82,0.11,1,0.25}
\definecolor{myyellow}{RGB}{0.4,0.8,0.1}

\begin{document}
\title{Visible shapes of black holes M87* and SgrA*} 
\author{V. I. Dokuchaev}\thanks{dokuchaev@inr.ac.ru}
\affiliation{Institute for Nuclear Research of the Russian Academy of Sciences, prospekt 60-letiya Oktyabrya 7a, Moscow 117312, Russia}
\author{N. O. Nazarova}\thanks{nnazarov@sissa.it}
\affiliation{Scuola Internazionale Superiore di Studi Avanzati (SISSA), Via Bonomea 265, 34136 Trieste (TS) Italy}
\affiliation{International Centre for Theoretical Physics (ICTP), Strada Costiera 11, 34151 Trieste (TS) Italy}

\date{\today}

\begin{abstract}
We review the physical origins for possible visible images of the supermassive black hole M87* in the galaxy M87 and SgrA* in the Milky Way Galaxy. The classical dark black hole shadow of the maximal size is visible in the case of luminous background behind the black hole at the distance exceeding the so-called photon spheres. The notably smaller dark shadow (dark silhouette) of the black hole event horizon is visible if the black hole is highlighted by the inner parts of the luminous accreting matter inside the photon spheres. The first image of the supermassive black hole M87 *, obtained by the Event Horizon Telescope collaboration, shows the lensed dark image of the southern hemisphere of the black hole event horizon globe, highlighted by accreting matter, while the classical black hole shadow is invisible at all. A size of the dark spot on the Event Horizon Telescope (EHT) image agrees with a corresponding size of the dark event horizon silhouette in a thin accretion disk model in the case of either the high or moderate value of the black hole spin, $a\gtrsim0.75$.
\end{abstract}
\keywords{General relativity \and Black holes \and Event horizon \and Gravitational lensing}
\maketitle \tableofcontents

\section{Introduction}

The enigmatic black holes are really black objects in the sky due to their physical properties. The famous quantum Hawking thermal radiation of black holes can be neglected in the case of numerous astrophysical black holes originated from the gravitational collapse of old massive stars. Nowadays, the only way to view black holes in the sky is a watching of black hole candidates highlighted by the surrounding matter. General relativity (Einstein's theory of gravity) predicts the appearance of dark black hole images on the surrounding luminous background. The Event Horizon Telescope (EHT) collaboration presents the first image of the supermassive black hole M87* at 1.3 mm wavelength with an unprecedented high angular resolution \cite{EHT1,EHT2,EHT3,EHT4,EHT5,EHT6}. Indeed, this image is the direct experimental evidence of the black hole existence in the Universe besides the famous observations of gravitational waves from the coalescence of black holes by the LIGO collaboration.

The visible shapes of black hole images depend on the distribution of emitting matter around black holes. We describe below the possible visible shapes of black hole images of the central supermassive black hole M87* in the galaxy M87 and SgrA* in our native Milky Way Galaxy. 

We show that the unique physical properties of the Kerr metric for rotating black hole \cite{Kerr} provide two qualitatively different forms of the visible black images: the standard black hole shadow or the notably smaller event horizon shadow (or event horizon silhouette). A particular visible black image crucially depends on the prevalence of emitting matter outside or inside of the so-called ``{\sl photon spheres}'' (see details and properties of photon spheres in Chapter~\ref{PhotonSpheres}).  

A standard black hole shadow is visible in the case of emitting matter placed outside the photon spheres (e.\,g., if there is a distant luminous background of extended hot gas clouds or luminous stars far outside the black hole). Meantime, a notably smaller event horizon silhouette is viewed, which is a shadow of the event horizon itself, in the case of emitting matter placed inside the photon spheres (e.\,g., if there is a highly luminous accreting matter in the vicinity of event horizon). To distinguish these two different black hole images, we will use in the following the term  {\sl ''classical black hole shadow''} for black hole image in the case of emitting matter placed outside the photon spheres (e.\,g., if there is a luminous stationary background far behind the black hole). 

We demonstrate below that on the first image of the supermassive black hole M87*, obtained by the Event Horizon Telescope collaboration, it is viewed namely the lensed dark image of the southern hemisphere of the black hole event horizon globe, highlighted by an accretion disk, while the classical black hole shadow is invisible at all.

The major scientific goal of the EHT collaboration is the registration of the supermassive black hole SgrA* image at the center of Milky Way \cite{Fish16,Lacroix13,Kamruddin,Johannsen16,Johannsen16b,Broderick16,Chael16,Kim16,Roelofs17,Doeleman17}. This supermassive black hole is the nearest ``dormant'' or ``sleeping'' quasar with a very low radiation activity and the mass $M=(4.3\pm0.3)\times10^6M_\odot$ \cite{Gheetal08,Gillessen09,Giletal09-2,Meyer12,Johannsen12}. Our native supermassive black hole SgrA* is evidently an object of intensive investigations \cite{Baade46,Becklin68,Eckart96,Dokuchaev77,Dokuchaev89,Allen90,Dokuchaev91,Dokuchaev91b,Manko92,Lo93,Backer93,Haller96,Ghez98,Backer99,Reid99,Baganoff99,Falcke00b,NovikovFrolov01,Baganoff01,Hornstein02,Genzel03,Aschenbach04,YusefZadeh06,Marrone08,Ghez08,Doeleman08,Doeleman09,DoddsEden09,Broderick09,Sabha10,Dexter10,Paolis11,Broderick11,Neilsen13,Zakharov13,Fish14,Gwinn14,Johnson14,Dokuch14,Moscibrodzka14,Bower15,Johnson15,Chatzopoulos15,FizLab,Rauch16,Zakharov16,Becerril16,Giddings16,Johannsen16c,OrtizLeon16,Parsa17,Capellupo17,Shiokawa17,Johnson17,Eckart17,Abdujabbarov17,Ponti17,Zajacek18,Abuter18,Zakharov18a,Zakharov18b,Zakharov18c,Zhu19,Izmailov19,Zakharov19,TuanDo19,Do19,Giddings19,Dai19,Moriyama19}. The other goal of the EHT collaboration is the registration of the supermassive black hole M87* with the mass $M=(6.6\pm0.4)\times10^9M_\odot$ in the nearest to us giant elliptical galaxy M87 (NGC 4486), which is placed in the central part of the Virgo cluster of galaxies \cite{Ho08,Gebhardt09,Gebhardt11,Walsh13}. The technological levels of the EHT and similar projects BlackHoleCam \cite{Goddi17} and GRAVITY \cite{GRAVITY18,GRAVITY19} permit to reach the angular resolution that matches the event horizon size of these supermassive black holes and allow to get the black hole image 
\cite{Mielnik62,Synge63,Bardeen73,Young76,Chandra,Falcke00,Takahashi04,Kardashev07,Falcke13,Li14,Inoue14,Cunha15,Abdujabbarov15,Younsi16}. 

The construction of advanced EHT version opens the new stage of the investigation of dark black hole silhouettes, as well as the testing general relativity and modified gravitation theories in the strong field limit \cite{deVries00,Schnittman06,Shatskiy08,Bambi09,Frolov09,Tamburini11,Vincent11,Amarilla12,Johannsen13,Babichev13,Amarilla13,Zakharov14,Wei15,Abd15,Nucamendi15,Nucamendi16,Cunha16,Abdujabbarov16,CliffWill17a,Cunha17,CliffWill17b,Amarilla17,Mureika17,DokEr15,Amir18,Lan18,Wang18,Lamy18,Mizuno18,Repin18,Hennigar18,Wei19,Blackburn19,Meierovich19,Abdikamalov19,Zhu19b,Tian19,Davoudiasl19,Konoplya19b,Hess19,Rummel20,Alexeyev20,Chagoya20}. The promising breakthrough for similar future investigations would be a construction of cosmic  interferometer with a nanosecond angular resolution \cite{Kardashev14,Roelofs19,Palumbo19}. 

In standard astrophysical conditions, the brightness of the accreting disk greatly exceeds the corresponding one of the distant luminous background, consisting of the extended hot gas clouds and bright stars. For this reason, the classical black hole shadow is complicated to observe in comparison with the event horizon silhouette (the shadow of the event horizon itself).

In Section \ref{classical} we describe the general properties of the classical black hole shadow, when the black hole is highlighted by a distant luminous background. In Section~\ref{PhotonSpheres} we elucidate the principal properties of photon spheres, which are crucial for understanding the possible forms of black hole images. At last, in the most original Section~\ref{silhouette} we explain the significant features of the event horizon silhouette (the event horizon shadow), produced by photons from the highly luminous accretion disk. 

The line element of the classical Kerr metric \cite{Kerr,Chandra,BoyerLindquist,Carter68,deFelice,Bardeen70,Bardeen70b,BPT,mtw,Galtsov}, describing the rotating black hole in standard Boyer--Lindquist coordinates $(t,r,\theta,\phi)$ \cite{BoyerLindquist}, is 
\begin{equation}
	ds^2=-e^{2\nu}dt^2+e^{2\psi}(d\phi-\omega dt)^2 +e^{2\mu_1}dr^2+e^{2\mu_2}d\theta^2,
	\label{metric}
\end{equation}
where
\begin{eqnarray}
	e^{2\nu}&=&\frac{\Sigma\Delta}{A}, \quad e^{2\psi}=\frac{A\sin^2\theta}{\Sigma}, 
	\quad e^{2\mu_1}=\frac{\Sigma}{\Delta}, \quad e^{2\mu_2}=\Sigma, \quad
	\omega=\frac{2Mar}{A}, \label{omega} \\
	\Delta &= & r^2-2Mr+a^2, \quad \Sigma=r^2+a^2\cos^2\theta, \quad A=(r^2+a^2)^2-a^2\Delta\sin^2\theta. \label{A}
\end{eqnarray}
In these equations $M$ --- black hole mass, $a=J/M$ --- black hole specific angular momentum (spin), $\omega$ --- frame-dragging angular velocity. We use the units with the gravitational constant $G=1$ and the velocity of light $c=1$. For simplification of formulas in the following, we often use the dimensional values for space distances $r\Rightarrow r/M$, for time intervals $t\Rightarrow t/M$ and etc. In other words, we will measure the radial distances in units $GM/c^2$ and time intervals in units $GM/c^3$. We also will use the dimensionless value for black hole spin $a=J/M^2\leq1$, by supposing that $0\leq a\leq1$. The black hole event horizon  radius $r_{\rm h}$ is the largest root of the quadratic equation $\Delta=0$:
\begin{equation}
	r_{\rm h}=1+\sqrt{1-a^2},
	\label{rh}
\end{equation}
There are four integrals of motion for test particles in the Kerr metric: $\mu$ --- test particle mass, $E$ --- particle total energy, $L$ --- particle azimuth angular momentum and $Q$ --- Carter constant, related with the non-azimuth angular momentum of the test particle and with non-equatorial motion \cite{Carter68}. The corresponding first-order differential equations of motion for test particle are \cite{Carter68,deFelice,Chandra,BPT,Bardeen70,Bardeen70b,mtw,Galtsov}:
\begin{eqnarray} \label{A9}
	\Sigma\frac{dr}{d\tau} &=& \pm \sqrt{R(r)}, \label{rmot} \\
	\label{A10}
	\Sigma\frac{d\theta}{d\tau} &=& \pm\sqrt{\Theta(\theta)}, \label{thetamot} \\
	\label{A11}
	\Sigma\frac{d\phi}{d\tau} &=& L\sin^{-2}\theta+a(\Delta^{-1}P-E), \\
	\label{A12}
	\Sigma\frac{dt}{d\tau} &=& a(L-aE\sin^{2}\theta)+(r^2+a^2)\Delta^{-1}P.
\end{eqnarray}
Here $\tau$ --- the proper particle time or affine parameter along the trajectory of massless ($\mu=0$) particle. In these equations the effective radial potential $R(r)$ governs the radial motion in these equations:
\begin{equation}
	R(r) = P^2-\Delta[\mu^2r^2+(L-aE)^2+Q],
	\label{Rr} 
\end{equation}
where $P=E(r^2+a^2)-a L$, and, respectively, the effective polar potential $\Theta(\theta)$ defines the polar motion of test particles:
\begin{equation}
	\Theta(\theta) = Q-\cos^2\theta[a^2(\mu^2-E^2)+L^2\sin^{-2}\theta].
	\label{Vtheta} 
\end{equation}
In particular, the zeros of these potentials define the turning points $dR/d\tau=0$ and $d\Theta/d\tau=0$ in the radial and polar directions, respectively. 

All trajectories of massive test particles ($\mu\neq0$) depend on three parameters (constants of motion or orbital parameters): $\gamma=E/\mu$, $\lambda=L/E$ and $q=\sqrt{Q}/E$. Respectively, the corresponding trajectories of massless particles ($\mu\neq0$) depends only on two parameters: $\lambda=L/E$ and $q=\sqrt{Q}/E$. From equations of motion (\ref{A9})--(\ref{A12}) it follows that Carter constant $Q\geq0$ for all particle trajectories reaching the space infinity at $r=\infty$. At finite distances from Kerr black hole there are ``{\sl vortex}'' orbits of test particles with $Q<0$ \cite{Wilkins}. The vortex orbits are beyond the scope of this article because we are interesting mainly the photon trajectories with $Q\geq0$, reaching a distant observer very far from the black hole, formally at $r=\infty$.

There are the integral form  \cite{Carter68,BPT,Chandra,mtw} of equations of test particle motion in the Kerr Metric, which are useful for numerical calculations:
\begin{equation}\label{eq2425a}
	\fint\frac{dr}{\sqrt{R(r)}}
	=\fint\frac{d\theta}{\sqrt{\Theta(\theta)}}, 
\end{equation}
\begin{equation}\label{eq2425b}
	\tau=\fint\frac{r^2}{\sqrt{R(r)}}\,dr
	+\fint\frac{a^2\cos^2\theta}{\sqrt{\Theta(\theta)}}\,d\theta,
\end{equation}
\begin{equation}\label{eq25ttc}
	\phi=\fint\frac{aP}{\Delta\sqrt{R(r)}}\,dr
	+\fint\frac{L-aE\sin^2\theta}{\sin^2\theta\sqrt{\Theta(\theta)}}\,d\theta, 
\end{equation}
\begin{equation}\label{eq25ttd}
	t=\fint\frac{(r^2+a^2)P}{\Delta\sqrt{R(r)}}\,dr
	+\fint\frac{(L-aE\sin^2\theta)a}{\sqrt{\Theta(\theta)}}\,d\theta. 
\end{equation}
In these equations the effective potentials $R(r)$ and $\Theta(\theta)$ are defined in equations (\ref{Rr}) and (\ref{Vtheta}). These specific integrals in (\ref{eq2425a})--(\ref{eq25ttd}) are the contour (path) integrals along the particle trajectory. These contour integrals are monotonic growing along the particle trajectory: the integrands in these contour integrals do not change their signs in transition through the radial and polar  turning points. In particular, the contour integrals along a particle trajectory in (\ref{eq2425a}) come to the ordinary ones if there are no radial and polar turning points along the particle trajectory
\begin{equation}\label{eq24a}
	\int^{r_s}_{r_0}\frac{dr}{\sqrt{R(r)}}
	=\int_{\theta_0}^{\theta_s}\frac{d\theta}{\sqrt{\Theta(\theta)}},
\end{equation}
where $r_s$ and $\theta_s$ --- the initial (starting) particle radial and polar angle coordinates. Respectively, in the case of a trajectory with only one turning point $\theta_{\rm min}(\lambda,q)$ (the extreme point in the polar effective potential $\Theta(\theta)$), the contour integrals in (\ref{eq2425a}) are written through the ordinary integrals in the form
\begin{equation}\label{eq24b}
	\int_{r_s}^{r_0}\frac{dr}{\sqrt{R(r)}}
	=\int_{\theta_{\rm min}}^{\theta_s}\frac{d\theta}{\sqrt{\Theta(\theta)}}
	+\int_{\theta_{\rm min}}^{\theta_0}\frac{d\theta}{\sqrt{\Theta(\theta)}}.
\end{equation}
Also, the contour integrals in (\ref{eq2425a}) in the case of the trajectory with two turning points $\theta_{\rm min}(\lambda,q)$ and $r_{\rm min}(\lambda,q)$ (the extreme point in the radial effective potential $R(r)$), are written through the ordinary integrals in the form
\begin{equation}\label{eq24c}
	\int_{r_{\rm min}}^{r_s}\frac{dr}{\sqrt{R(r)}}
	+\int_{r_{\rm min}}^{r_0}\frac{dr}{\sqrt{R(r)}}
	=\int_{\theta_{\rm min}}^{\theta_s}\frac{d\theta}{\sqrt{\Theta(\theta)}}
	+\int_{\theta_{\rm min}}^{\theta_0}\frac{d\theta}{\sqrt{\Theta(\theta)}}.
\end{equation}

\section{Classical black hole shadow: black hole highlighting by distant luminous background}
\label{classical}

The classical black hole shadow is a capture photon cross-section in the black hole gravitational field. It is observable if there is a distant luminous background behind the black hole at the distance, exceeding the corresponding radius of the photon spheres (see definition and specific features of photon spheres in Section~\ref{PhotonSpheres}). The classical black hole shadow is investigated in details in numerous works \cite{Mielnik62,Synge63,Bardeen73,Young76,Chandra,Falcke00,Luminet79,Zakharov94,Beckwith05,ZakhPaoIngrNuc05,Takahashi05,Takahashi07,Bakala07,Huang07,Virbhadra09,Hioki09,Schee09,Dexter09,JohPsaltis10,Amarilla10,Nitta11,Yumoto12,Abdujabbarov13,Bambi13,Atamurotov13,Atamurotov13b,Wei13,Tsukamoto14,Papnoi14,Tinchev14,Kraniotis14,Ghas15,Tinchev15,Gralla15,Atamurotov15,Perlick15,Shipley16,Liu16,Yang16,Strom16,Amir16,Gralla16,Vincent16,Dastan16,Tret16,Dastan16b,Sharif16,Opatrny17,Cunha17b,Singh17,Wang17,Amir17,Strom17,Strom18,Tsupko17a,Tsupko17b,Cunha18b,BisnovatyiTsupko17,Stuchlik18,Huang18,Tsukamoto18,Bisnovatyi18,Hou18,Yan19,Vagnozzi19,Gyulchev19,Kumar19,Konoplya19,Sabir19,Johnson19,Siino19,Zhang19,Shipley19,Shaikh19,Shaikh19b,Ding19,Narayan19,Goddi19,Feng19,Allahyari19,Konoplya19c,doknazsm19,dokuch19,Cunha20,Tsupko20,Vagnozzi20,Yu20,Banerjee20,gralla20,Chang20,Himwich20,Li20,Jusufi20,Bakala20,Anantua20,Belhaj20}.

The observed outline (contour) of the classical black hole shadow, projected on the celestial sphere, is defined by the photon orbits with a constant radius, $r=r_{\rm ph}=const$, named either the spherical photon orbits or photon spheres. In a general case of the rotating Kerr black hole (with a black hole spin $a\neq0$) the photons on spherical orbits are moving in the azimuth and polar (latitude) directions on the surface of a constant radius $r=r_{\rm ph}$ by oscillating in the polar direction between the minimum $\theta_{\rm min}$ and maximum $\theta_{\rm max}=\pi-\theta_{\rm min}$ polar angles (see. definitions and details in Chapter~\ref{PhotonSpheres}).

\begin{figure}
	\centering
	\includegraphics[angle=0,width=0.7\textwidth]{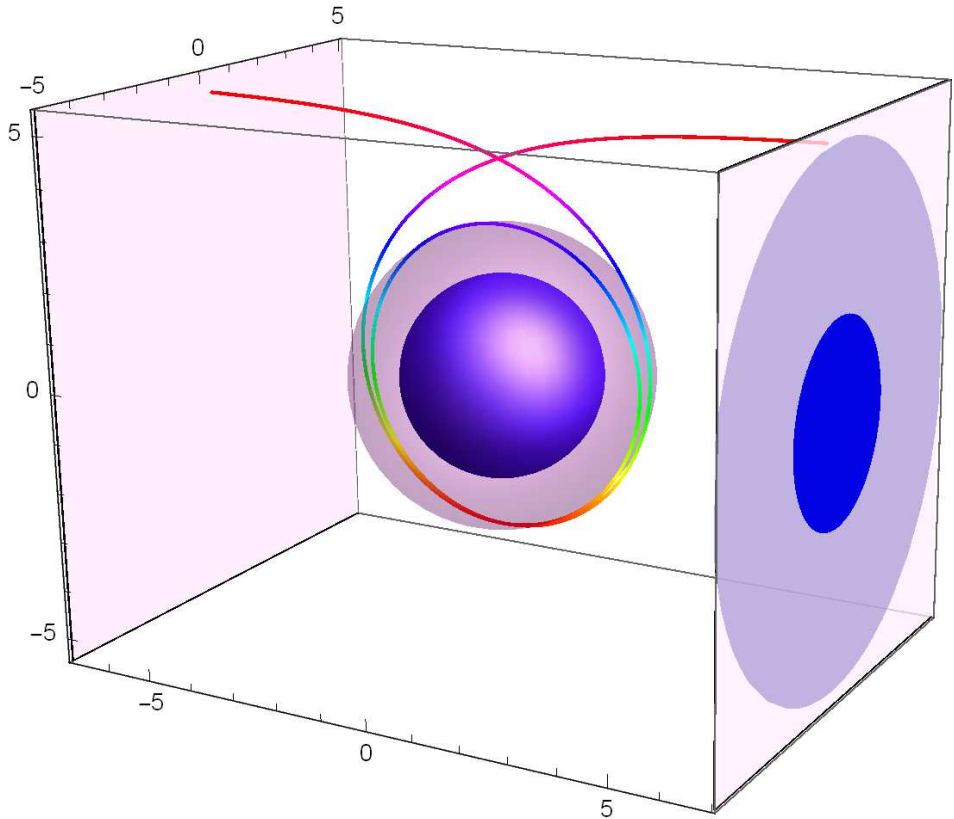} 
	\caption{The classical shadow (lighter {\color{blue}blue} disk) of the Schwarzschild black hole ($a=0$), highlighted by a distant luminous background. It is shown a typical $3D$ photon trajectory (multicolored $3D$ curve), which is starting from the distant background. Then, this trajectory is winding multiply near the radius of circular photon orbit around the black hole event horizon globe (darker {\color{blue}blue} sphere) at the return radius $r_{\rm min}=r_{\rm ph}=3$. The finishing point of this photon trajectory is the north polar point on the outline of the black hole shadow, viewed by a distant observer. Inside the black hole shadow is shown a fictitious image (dark {\color{blue}blue} disk with a radius $r=2$) of the lensed image of the black hole event horizon in the imaginary Euclidean space (in the absence of gravity). The circular orbits of photons, producing the outline of shadow, are placed on the {\color{purple}purple} photon sphere with the radius $r=3$.}
	\label{fig1}
\end{figure}

The shape of a classical black hole shadow in the Kerr metric is defined analytically in the parametric form $(\lambda,q)=(\lambda(r),q(r))$, namely (see, e.\,g., \cite{Bardeen73,Chandra}):
\begin{equation}
	\lambda=\frac{(3-r)r^2-a^2(r+1)}{a(r-1)}, \quad
	q^2=\frac{r^3[4a^2-r(r-3)^2]}{a^2(r-1)^2},
	\label{shadow}
\end{equation}
where $r$ is a radius of photon sphere and $\lambda$ and $q$ are, respectively, the horizontal and vertical impact parameters of photons on the celestial sphere, viewed by a static distant observer in the black hole equatorial plane. In this equation $r$ is a radius of the photon sphere for given impact parameters $\lambda$ and $q$. Strictly speaking, equations (\ref{shadow}) with $q\geq0$ reproduce only upper half of the shadow. Lower half of shadow is a mirror reflection of the upper one with respect to the black hole equatorial plane (due to the Kerr metric reflection symmetry over the equatorial plane). 

James Bardeen named the classical black hole shadow of the Kerr black hole as the ``viewed boundary'' of the black hole in his pioneering work  \cite{Bardeen73}. For a more general modern definition of the classical black hole shadow see, e.\,g., \cite{Grenzebach14,Grenzebach15,Cunha18a}.

\begin{figure}[t]
	\centering
	\includegraphics[angle=0,width=0.75\textwidth]{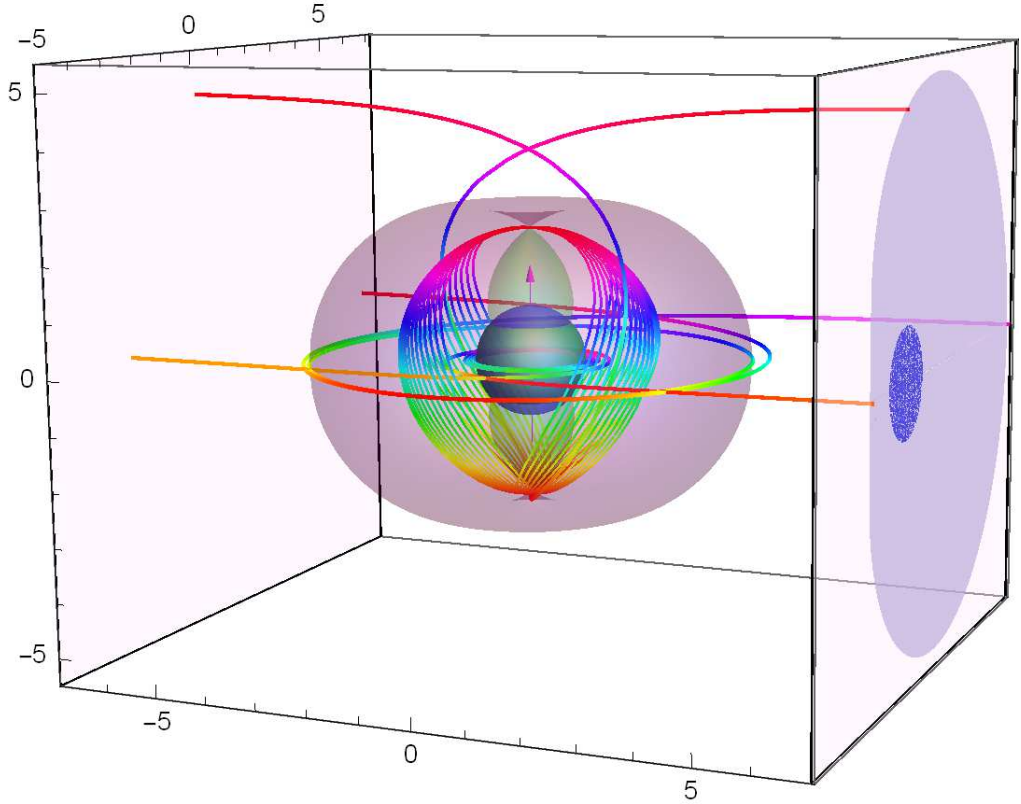} 
	\caption{The classical shadow  (closed magenta region) of the extreme Kerr black hole ($a=1$), highlighted by the distant luminous background and viewed by a distant observer in the black hole equatorial plane. A closed {\color{purple}purple} region is an envelope of photon spheres for photons with $\lambda\leq0$, while a closed {\color{mygreen}green} region is an envelope of photon spheres for photons with $\lambda\geq0$. Multicolored curves are the examples of photon trajectories, producing an outline (boundary) of the classical black hole shadow, with orbital parameters, respectively, $(\lambda,q)=(2,0)$ --- corotating skimming photon in the equatorial plane, $(\lambda,q)=(-7,0)$ --- counter-rotating photon in the equatorial plane and $(\lambda,q)= (0,\sqrt{11+8\sqrt{2}})$ --- fully spherical photon orbit. A blue disk with a radius $r=1$ inside the black hole shadow is the observed position of the black hole event horizon in the imaginary Euclidean space (in the absence of gravity). A magenta arrow is the black hole rotation axes.}
	\label{fig2}
\end{figure}

The photon spheres are reduced to photon circles with radius $r_{\rm ph}=3$ in the simplest limiting case of the Schwarzschild black hole ($a=0$). The corresponding radius of the classical black hole shadow in the Schwarzschild black hole case is $r_{\rm ph}=3\sqrt{3}$. 

Figure~\ref{fig1} shows the $3D$ illustration of the classical black hole shadow formation in the Schwarzschild black hole case ($a=0$), when the luminous background is placed  behind the black hole at the distance, exceeding the size of photon sphere ({\color{purple}purple} sphere with a radius $r_{\rm ph}=3$). Throughout this paper we use the standard (dimensionless) Boyer--Lindquist coordinates $(t,r,\theta\phi)$ in all $3D$ Figures similar to the Figure~\ref{fig1}.

In the limiting case of extreme Kerr black hole ($a=1$) the expressions (\ref{shadow}) for classical black hole shadow are simplified:
\begin{equation}
	\lambda=-r^2+2r+1, \quad q^2=r^3(4-r). \label{shadow3}
\end{equation}
It must be noted, that these limiting formulas do not produce the closed form of the outline (boundary) for the classical black hole shadow due to the nonuniform nature of the limit $a\to1$ in the Boyer-Lindquist coordinates. It must be added the vertical line $(r=1,0\leq q\leq\sqrt{3})$ to close the outline of shadow \cite{Bardeen73}. The inverted forms of expressions (\ref{shadow3}) are
\begin{equation}
	r_{\rm ph}(\lambda)=1+\sqrt{2-\lambda}, \quad q_{\rm ph}(\lambda)^2=(1+\sqrt{2-\lambda})^3(3-\sqrt{2-\lambda}). \label{shadow4}
\end{equation}

See in Figure~\ref{fig2} the corresponding $3D$ illustration of the classical black hole shadow formation in the extreme Kerr black hole case ($a=1$). As in Figure~\ref{fig1}, the luminous background is placed  behind the black hole at the distance, exceeding the size of all photon spheres. A closed {\color{purple}purple} region is an envelope of all photon spheres for photons with $\lambda<0$ and, respectively, a closed green region is an envelope of all photon spheres for photons with $\lambda>0$). Multicolored curves in this Figure are the examples of photon trajectories, producing an outline (boundary) of the classical black hole shadow, with orbital parameters, respectively, $(\lambda,q)=(2,0)$ --- corotating skimming photon in the equatorial plane, $(\lambda,q)=(-7,0)$ --- counter-rotating photon in the equatorial plane and $(\lambda,q)= (0,\sqrt{11+8\sqrt{2}})$ --- fully spherical photon orbit. A blue disk with a radius $r=1$ inside the black hole shadow is the observed position of the black hole event horizon in the imaginary Euclidean space (in the absence of gravity). A magenta arrow is the black hole rotation axes.

\begin{figure}[t] 
	\centering 
	\includegraphics[width=0.8\textwidth,origin=c,angle=0]{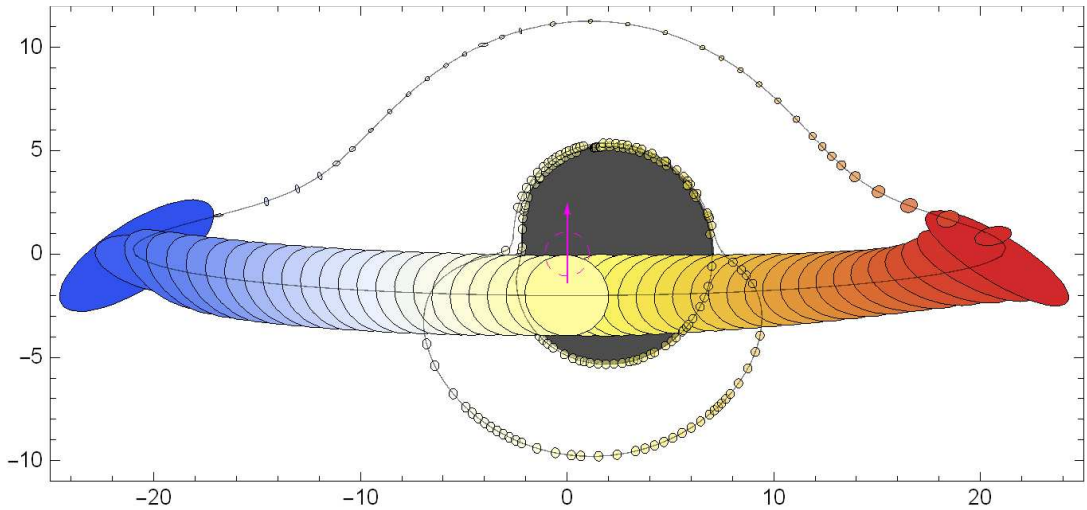}
	\caption{Direct images and also the first and second light echoes of the lensed compact star (luminous probe) in discrete times at the circular equatorial orbit with radius $r=20$ around a near extreme black hole SgrA* viewed by a distant observer. All multiple images of this star are placed outside the classical black hole shadow (the closed black region). A dashed {\color{magenta}magenta} circle is the projection of the black hole event horizon on the celestial sphere in the imaginary Euclidean space (in the absence of gravity).}
	\label{fig3}
\end{figure}

The gravitational lensing by black holes provides, in general, the infinite number of images \cite{CunnBardeen72,CunnBardeen73,Viergutz93,RauchBlandf94,GralHolzWald19}.
Christopher Cunningham and James Bardeen elaborated the very usable classification scheme for multiple images (or light echoes) \cite{CunnBardeen72,CunnBardeen73}, based on the number of intersections the black hole equatorial plane by photon on its way from the initial emission point to a distant observer. An astrophysical example of the stationary luminous background is shown in Figure~\ref{fig3}, demonstrating direct images and also the first and second light echoes of the lensed images of a compact star (luminous probe) in discrete times at the circular equatorial orbit with radius $r=20$ around a near extreme black hole. The orbital radius of this star exceeds the corresponding photon spheres. Therefore, this star plays a role of the distant stationary background and all its multiple images are placed outside the classical black hole shadow (see details in \cite{doknaz17} and animation of numerical modeling in \cite{doknaz18b}).

\section{Photon spheres}
\label{PhotonSpheres}

Among the others, a striking feature of the Kerr metric is the existence of relativistic spherical orbits for massive and massless particles moving on the sphere $r=const$ and oscillating in polar (latitude) direction between the turning points. From differential equations of motion in the Kerr metric (\ref{A9})--(\ref{A12}) it follows that parameters of spherical orbits are defined by the common solutions of equations $R(r)=dR(r)/dr=0$, where the radial effective potential $R(r)$ is from (\ref{Rr}). Correspondingly, the polar turning points $\theta_{\rm min}$ and $\theta_{\rm max}=\pi-\theta_{\rm min}$ are defined by zeros of effective polar potential $\Theta(\theta)$ from (\ref{Vtheta}). Spherical orbits were described in the pioneering work by Daniel Wilkins \cite{Wilkins} (see also \cite{Claudel01,Grossman12,Hod13,Liu19,Glampedakis19,Hughes19,Teo20}). The spherical orbits are reduced to the circular ones in the limiting case of equatorial orbits with $q=0$ \cite{BPT}. 

The spherical orbits of photons are naturally named as ``{\sl photon spheres}''. The outline (contour) of classical black hole shadow is defined namely by photon spheres according to expressions (\ref{shadow}). 

\begin{figure}
	\centering
	\includegraphics[width=0.75\textwidth]{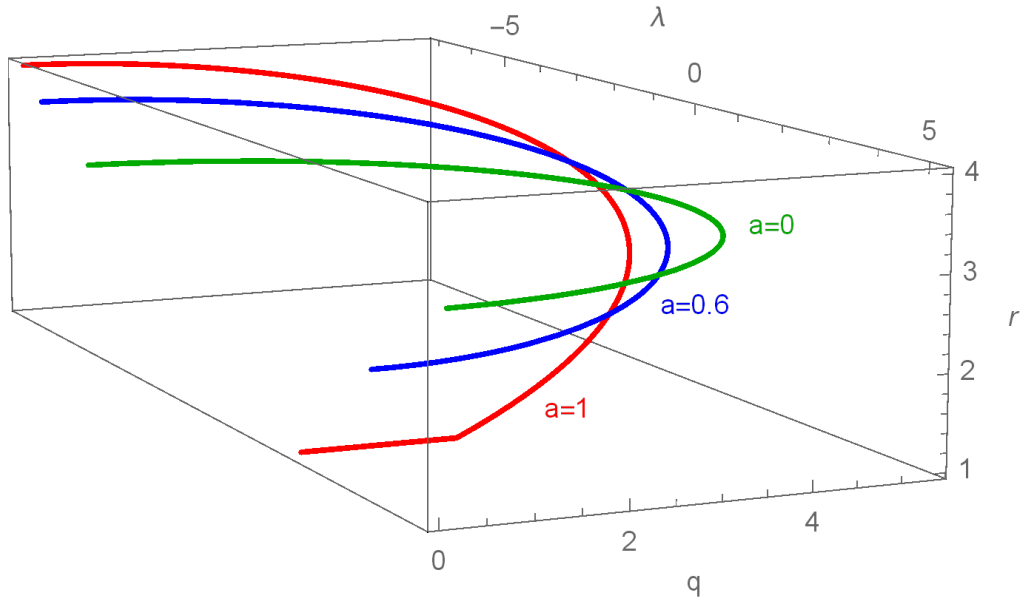}
	\caption{The radii of photon spheres depend on one of the photon orbit parameters, $\lambda$ or $q$. At $a=1$ the photon spheres ({\color{red} red} curve) are placed at radii $r_{\rm ph}=1+\sqrt{2-\lambda}$, $q^2= (1+\sqrt{2-\lambda})^3(3 -\sqrt{2-\lambda})$. The corresponding photon spheres exist at the radial interval $1\leq r_{\rm ph}\leq4$. At $a=0$ the photon sphere is reduced to the photon circle with radius ({\color{mygreen} green} semicircle) with $\lambda^2+q^2=27$. A {\color{blue}blue} curve corresponds to the radii of photon spheres at $a=0.6$.}
	\label{fig4}      
\end{figure}

Figure~\ref{fig4} shows the radii of photon spheres, depending on one of the photon orbit parameters $\lambda$ or $q$ according to expressions (\ref{shadow}). At $a=1$ the photon spheres ({\color{red} red} curve) are placed at radii $r_{\rm ph}=1+\sqrt{2-\lambda}$ at the radial interval $1\leq r_{\rm ph}\leq4$. At $a=0$ the photon sphere are reduced to photon circles with a radius $r_{\rm ph}=3$ ({\color{mygreen}green} semicircle) with $\lambda^2+q^2=27$. A {\color{blue}blue} curve corresponds to the radii of photon spheres at $a=0.6$.

A turning point in polar direction $\theta_{\rm min}$ on the spherical photon trajectory defined by the condition $\Theta(\theta)=0$. According to the Cunningham--Bardeen classification scheme of the multiple lensed images \cite{CunnBardeen72,CunnBardeen73}, the photons, providing the prime image of the emitting source, do not intersect the black hole equatorial plane on their way from the source to a distant observer. 

In the Schwarzschild black hole case ($a=0$) the corresponding turning point $\theta_{\rm min}= \arccos(q/(3\sqrt{3})$. In the Kerr case ($a\neq0$) the polar turning point (if it exists) is placed at 
\begin{equation}
	\cos^2\theta_{\rm min}=\frac{\sqrt{4a^2q^2+(q^2
			+ \lambda^2-a^2)^2}-(q^2+\lambda^2-a^2)}{2a^2}. 
	\label{thetamin} 
\end{equation} 
This expression for $\theta_{\rm min}$ is used in the numerical solution of integral equations (\ref{eq24a})--(\ref{eq24c}).

\begin{figure}
	\centering
	\includegraphics[angle=0,width=0.8\textwidth]{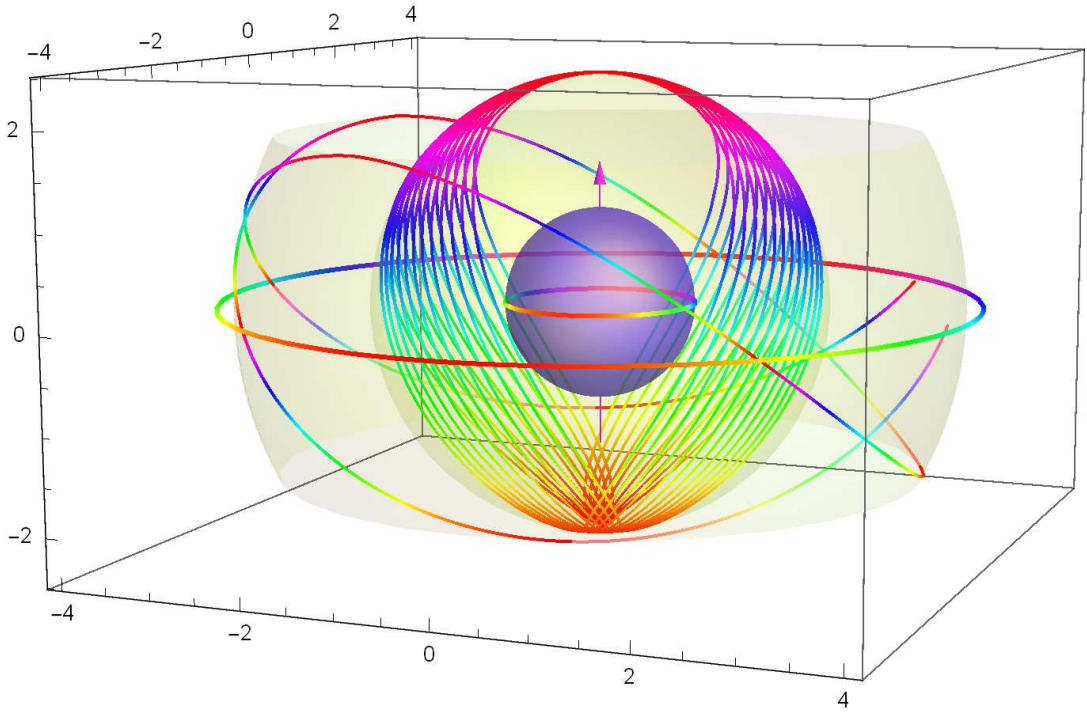}
	\caption{Examples of photon spheres in the case of the extreme Kerr black hole ($a=1$). The multicolored curves correspond to  spherical photon trajectories with orbital parameters, respectively, $(r,\lambda,q)=(1,2,0)$ --- co-rotating photon in the equatorial plane, $(r,\lambda,q)=(1+\sqrt{2},0,\sqrt{11+8\sqrt{2}})$ --- fully spherical photon orbit, $(r,\lambda,q)=(1+2\sqrt{2},-6,\sqrt{16\sqrt{2}-13})$ and $(r,\lambda,q)=(4,-7,0)$ --- counter-rotating photon in the equatorial plane.}
	\label{fig5}
\end{figure}

\begin{figure}
	\centering
	\includegraphics[width=0.4\textwidth]{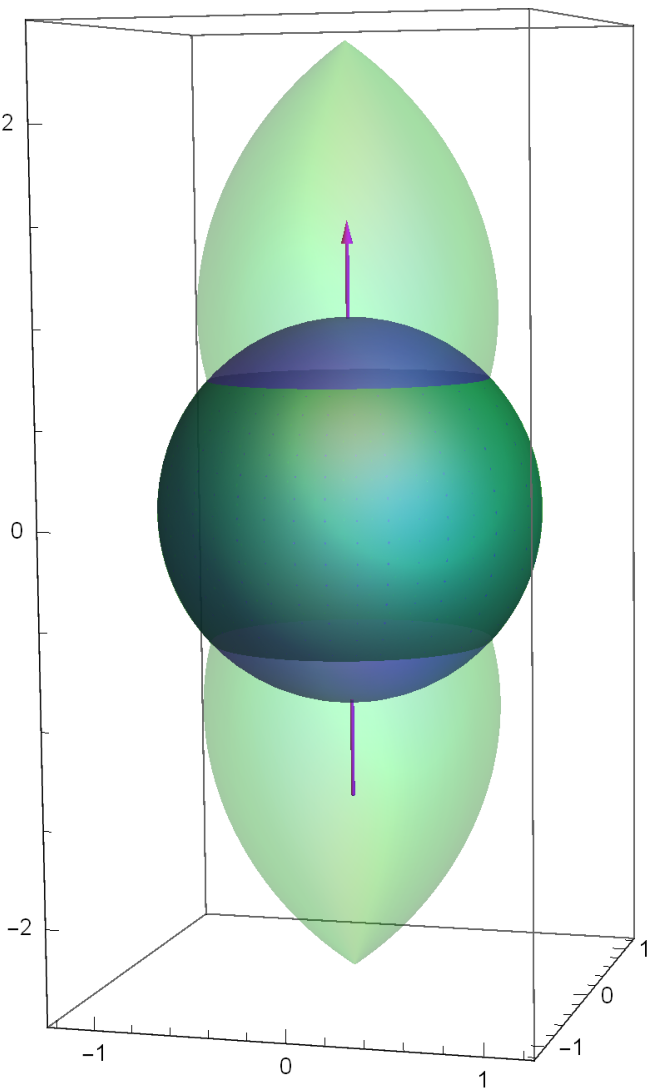}
	\includegraphics[width=0.45\textwidth]{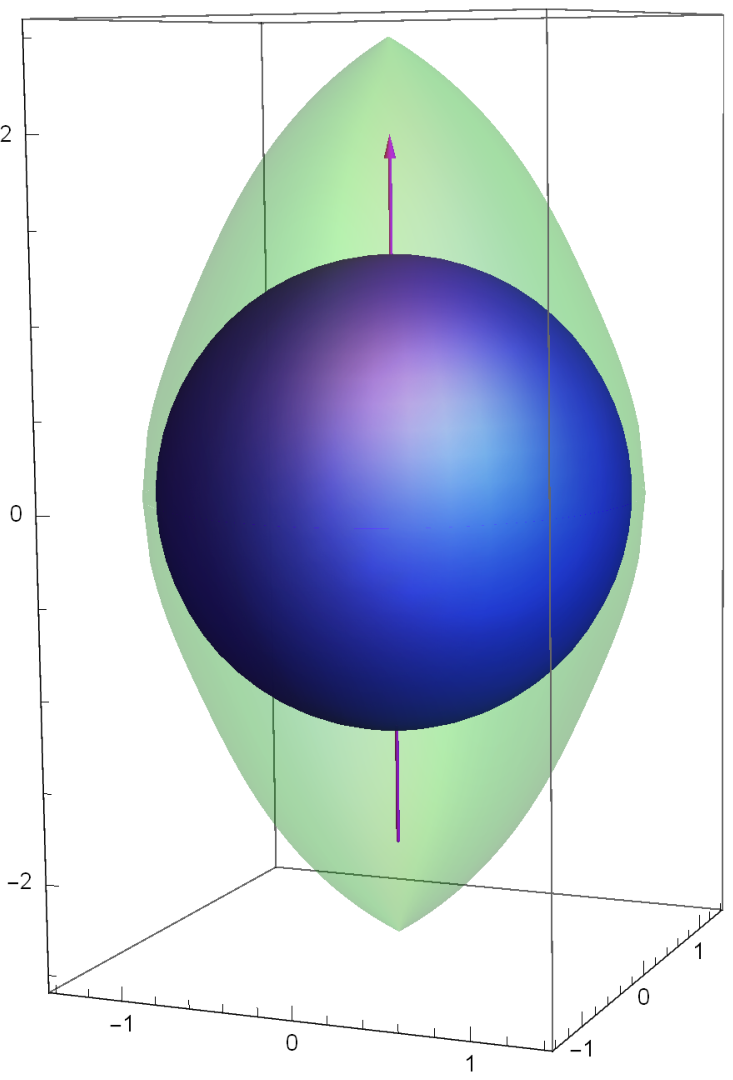}
	\caption{Envelopes of photon spheres with $\lambda\geq0$ (closed {\color{mygreen}green} regions) in the case of Kerr black holes with spin $a=1$ (left panel) and $a=0.95$ (right panel). In the case of extreme Kerr black hole ($a=1$) there are skimming photons, which are moving both in the azimuth and latitude directions on the sphere with radius $r=1$ by oscillating in polar directions with $0\leq\cos[\theta_{\rm min}]\leq\sqrt{2\sqrt{3}-3}$.}
	\label{fig6}
\end{figure}

Figure~\ref{fig5} shows some $3D$ examples of photon spheres (spherical photon trajectories) around the extreme Kerr black hole. These photon oscillates in polar direction between $\theta_{\rm min}$ and $\theta_{\rm max}= \pi-\theta_{\rm min}$. The multicolored curves in this Figure are the examples of spherical photon trajectories with orbital parameters, respectively, $(r,\lambda,q)=(1,2,0)$ --- co-rotating photon in the equatorial plane, $(r,\lambda,q)=(1+\sqrt{2},0,\sqrt{11+8\sqrt{2}})$ --- fully spherical photon orbit, $(r,\lambda,q)=(1+2\sqrt{2},-6,\sqrt{16\sqrt{2}-13})$ and $(r,\lambda,q)=(4,-7,0)$ --- counter-rotating photon in the equatorial plane.

Figure~\ref{fig6} shows $3D$ {\color{mygreen}green} regions, which are the envelopes of all photon spheres with with $\lambda\geq0$  in the case of Kerr black holes with spin $a=1$ and $a=0.95$. In the extreme black hole case with $a=1$ the green part of the event horizon globe is a region for the very specific photon spheres, which are called ``{\sl the skimming photons}'' \cite{Wilkins}. The orbit parameters of skimming photons are $\lambda=2$ and $0\leq q\leq\sqrt{3}$. These skimming photons move both in the azimuth and latitude direction on the sphere with radius $r=1$ by oscillating in polar directions with $0\leq\cos[\theta_{\rm min}]\leq\sqrt{2\sqrt{3}-3}$, as it follows from equations (\ref{shadow3}) and (\ref{thetamin}).

\section{Event horizon silhouette: black hole highlighting by accretion disk}
\label{silhouette}

In the general case of a static distant observer, placed at the given radius $r_0\gg r_{\rm h}$ ( e.\,g., practically at the space infinity), at the given polar angle $\theta_0$ and at the given azimuth $\phi_0$, it must be used the horizontal impact parameter $\alpha$ and vertical impact parameter $\beta$ on the celestial sphere (see details in \cite{Bardeen73,CunnBardeen72,CunnBardeen73}):
\begin{equation}
	\alpha =-\frac{\lambda}{\sin\theta_0}, \quad
	\beta = \pm\sqrt{\Theta(\theta_0)},
	\label{alpha} 
\end{equation}
where the effective polar potential  $\Theta(\theta)$ is from Equation (\ref{Vtheta}). 

In the simplest case of the spherically symmetric Schwarzschild black hole ($a=0$) the boundary of the event horizon image (the boundary of the dark event horizon silhouette), viewed by a distant observer (which is placed at $\theta_0=\pi/2$), is defined by solution of the integral equation
\begin{equation}
	\int_{2}^{\infty}\frac{dr}{\sqrt{R(r)}}
	=2\int_{\theta_{\rm min}}^{\pi/2}\frac{d\theta}{\sqrt{\Theta(\theta)}},
	\label{a0max}
\end{equation}
where $\theta_{\rm min}$ is a turning point in polar direction on the photon trajectory for the direct image of the small accreting fragment (probe), defined by the condition $\Theta(\theta)=0$. According to the Cunningham--Bardeen classification scheme of the multiple lensed images \cite{CunnBardeen72,CunnBardeen73}, the photons, providing the prime image of the emitting source do not intersect the black hole equatorial plane on their way from the source to a distant observer. The event horizon radius of the Schwarzschild black hole is $r_{\rm h}=2$, and turning point $\theta_{\rm min}= \arccos(q/\sqrt{q^2+\lambda^2})$. An integral in the right-hand-side of equation (\ref{a0max}) in this case is equal $\pi/\sqrt{q^2+\lambda^2}$. In result, the numerical solution of the integral equation (\ref{a0max}) gives for the radius of the event horizon image (silhouette) the value $r_{\rm eh}=\sqrt{q^2+\lambda^2}\simeq4.457$. This radius is notably smaller than the corresponding radius of the black hole shadow $r_{\rm sh}=3\sqrt{3}\simeq5.2$. 

The supermassive black hole SgrA* at the center of the Milky Way galaxy has a mass $M= (4.3\pm0.3)\times10^6M_\odot$, i.\,e., three orders of magnitude less than in the case of M87*, but at the same time the black hole SgrA* is placed three orders of magnitude closer than the black hole M87*. So the event horizons of these two black holes have approximately the same angular sizes accessible for observations by the EHT. Rotation axis orientation of the black hole SgrA* most probably coincides with the rotation axis of the Milky Way galaxy \cite{Psaltis15}. For concreteness, we suppose that distant observer is placed near the equatorial plane of the black hole SgrA* at $\cos\theta_0=0.1$ or $\theta_0\simeq84.24^\circ\!$.

\begin{figure}
	\centering	
	\includegraphics[width=0.7\textwidth]{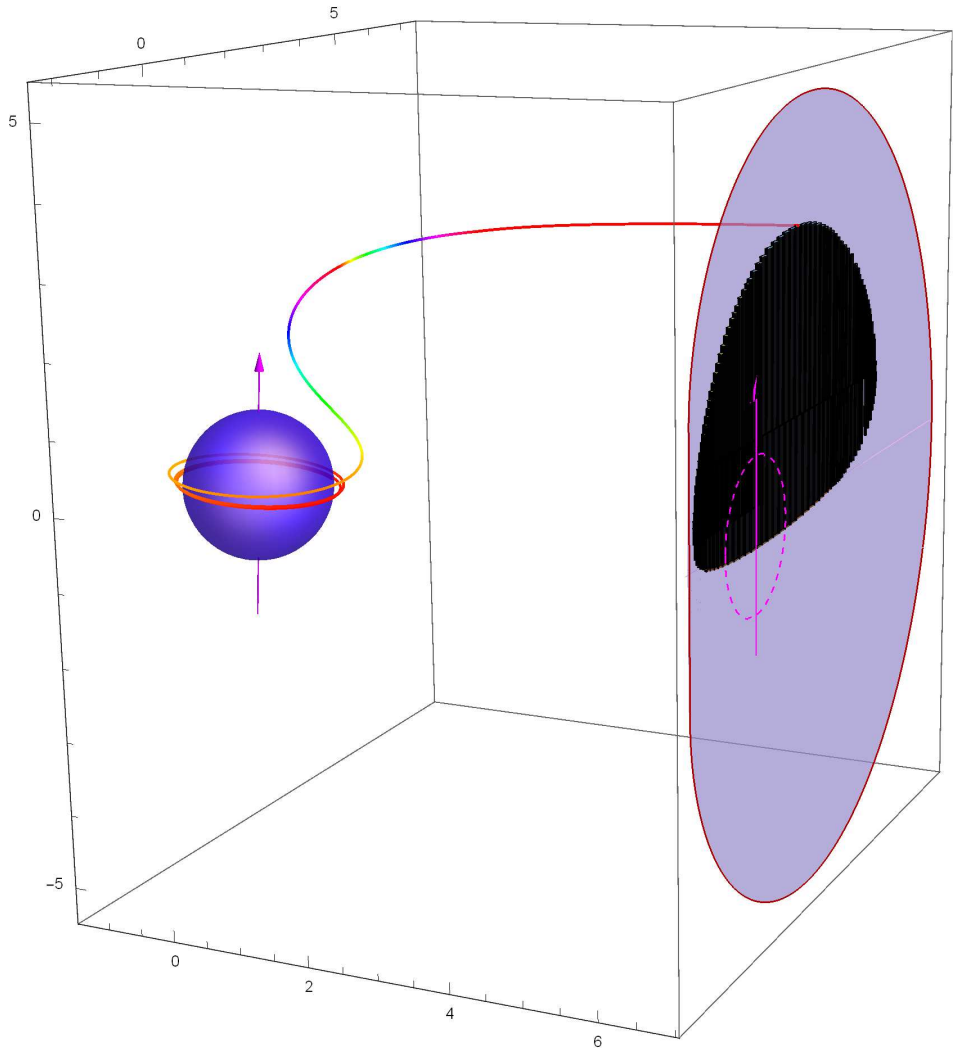}
	\caption{A visible dark silhouette of the northern hemisphere of event horizon (black region) illuminated by a thin accretion disk in the equatorial plane of the black hole with the spin $a=0.9982$, corresponding to the orientation of supermassive black hole SgrA* with respect to a distant observer. The outline (contour) of this silhouette is defined by the highly red-shifted photons, emitted near the black hole event horizon by the thin accreting disk and registered by a distant observer. It is shown the photon trajectory (multicolored $3D$ curve), producing the north pole point on the outline of the event horizon dark silhouette.}
	\label{fig7}      
\end{figure}

\begin{figure}
\centering	
\includegraphics[width=0.24\textwidth]{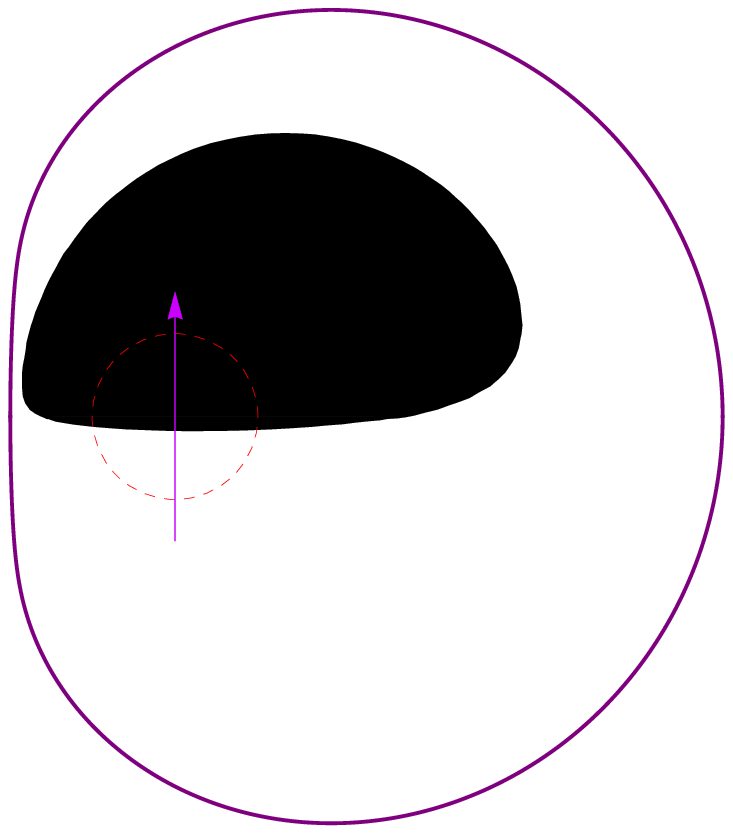}
\hskip0.1cm
\includegraphics[width=0.27\textwidth]{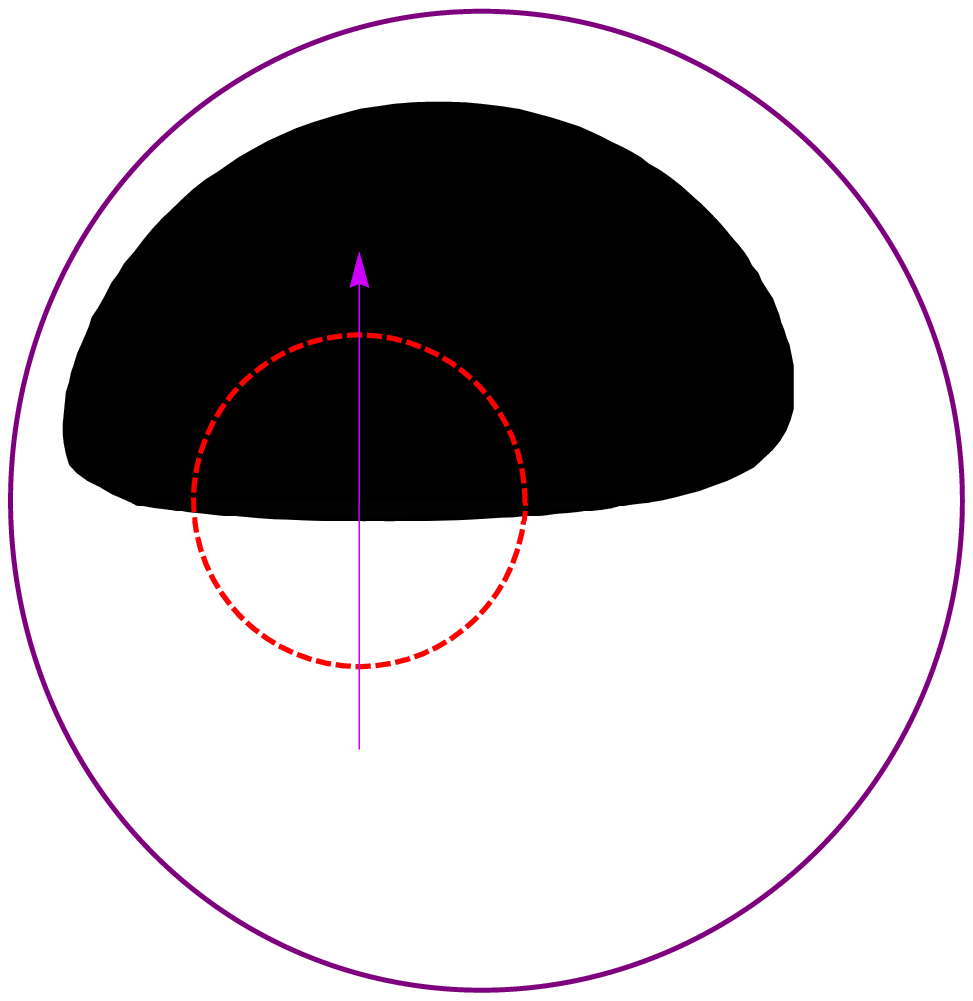}
\hskip0.1cm
\includegraphics[width=0.29\textwidth]{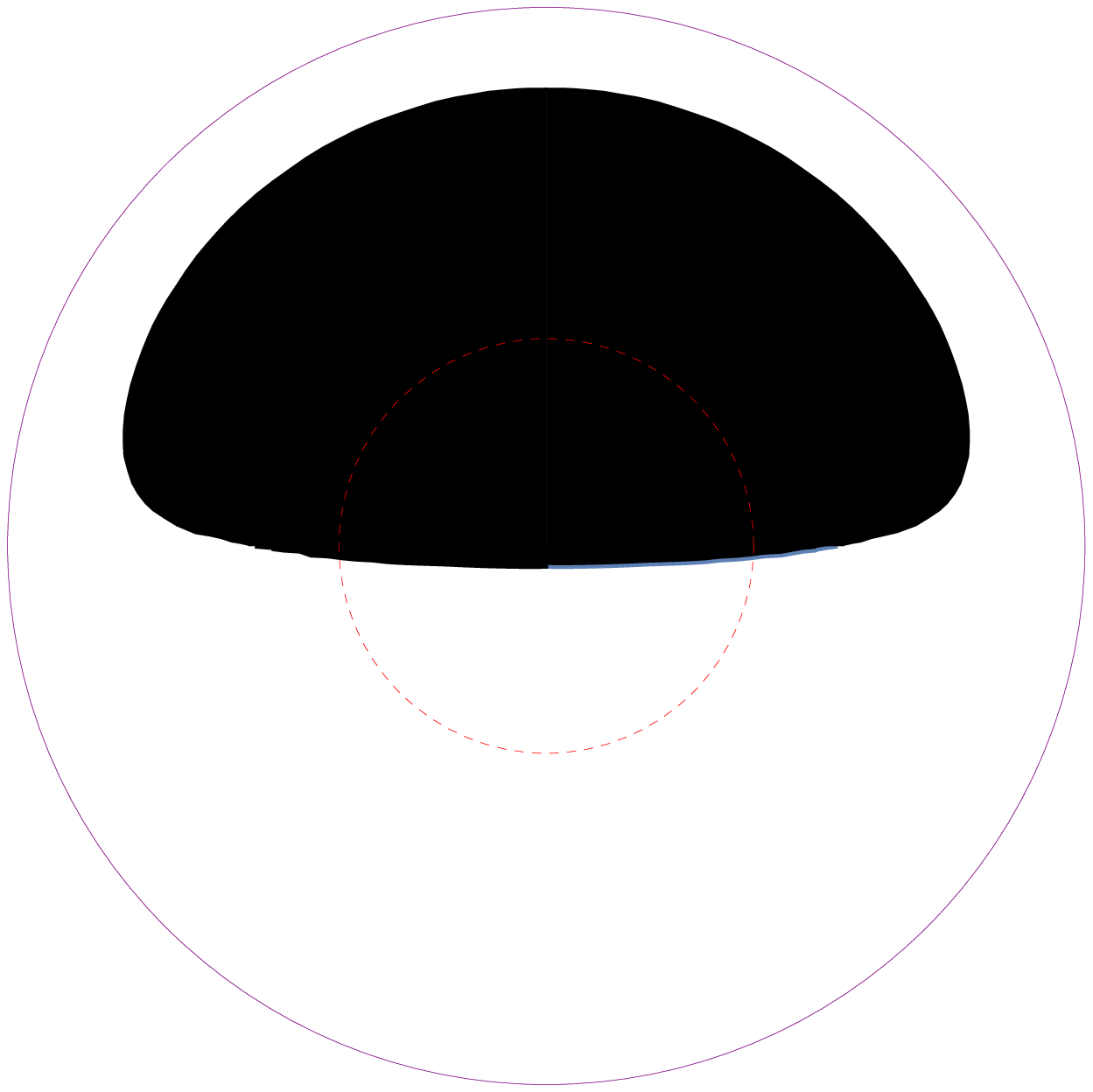}
\caption{The dark silhouettes of the northern hemisphere of the event horizon (black region) in the case of supermassive black hole SgrA* ($\theta_0=17^\circ$), projected inside the boundary of classical black hole shadow (closed {\color{purple}purple} curves) for the values of black hole spin, respectively, $a=0.9982$ (left), $0.65$ (middle) and $0$ (right).}
\label{fig8}      
\end{figure}


\begin{figure}
	\centering	
	\includegraphics[angle=0,width=0.7\textwidth]{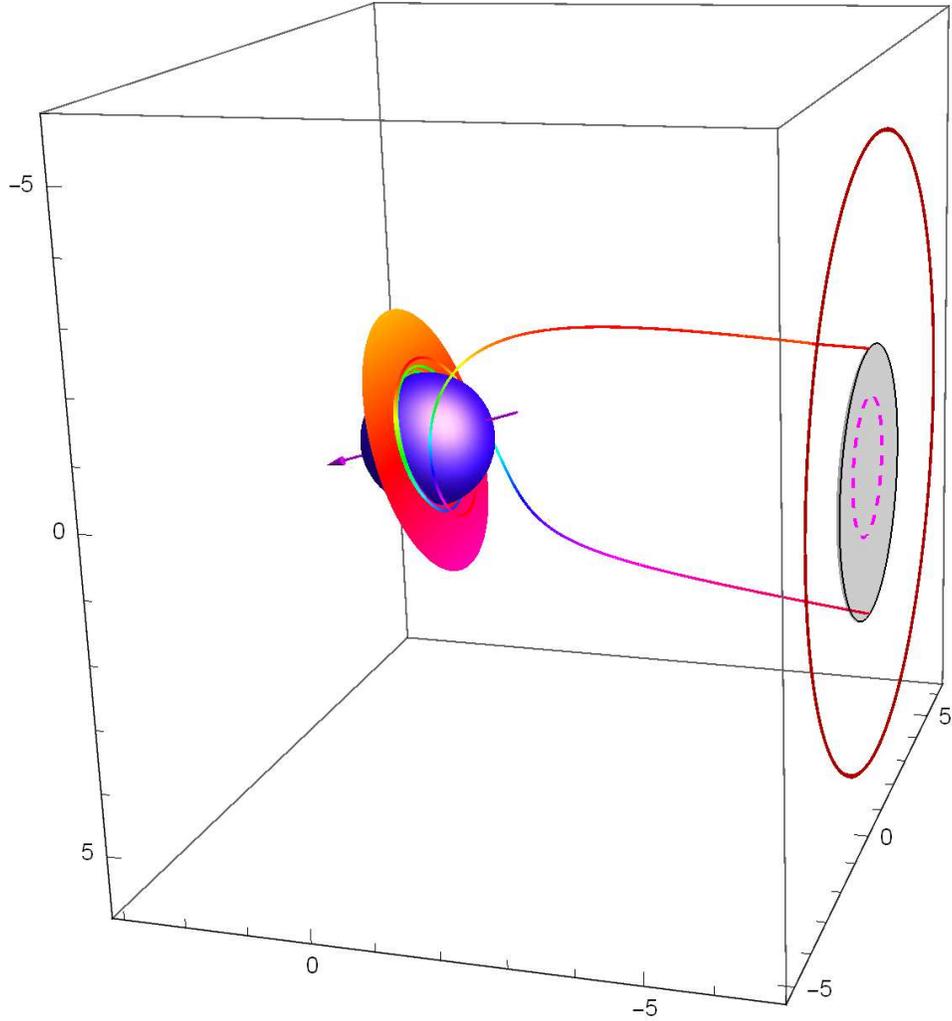} 
	\caption{Two $3D$ photon trajectories, starting from the different points of the circle with a radius $r=1.01\,r_{\rm h}$ at the thin accretion disk ({\color{orange}orange} oval)
		in the equatorial plane of rotating black hole with the spin $a=0.9982$ and reaching a distant observer near the outer contour of the event horizon silhouette (dark gray region). The closed  dark red curve is a projection of the classical black hole shadow contour on the celestial sphere.}
	\label{fig9}
\end{figure}

\begin{figure}
	\centering	
	\includegraphics[width=0.28\textwidth]{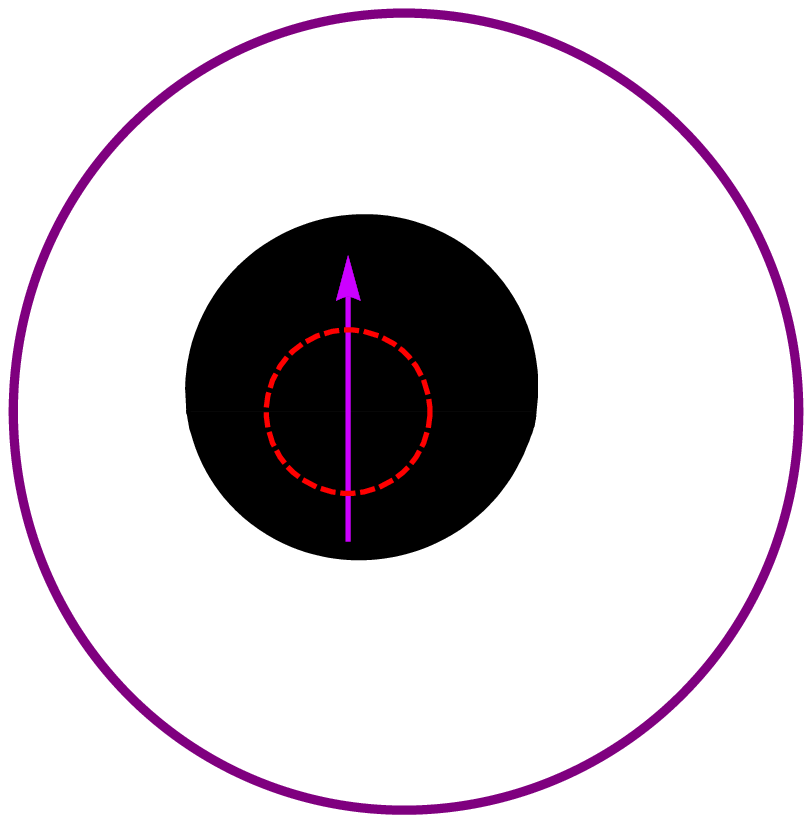}
	\hskip0.1cm
	\includegraphics[width=0.29\textwidth]{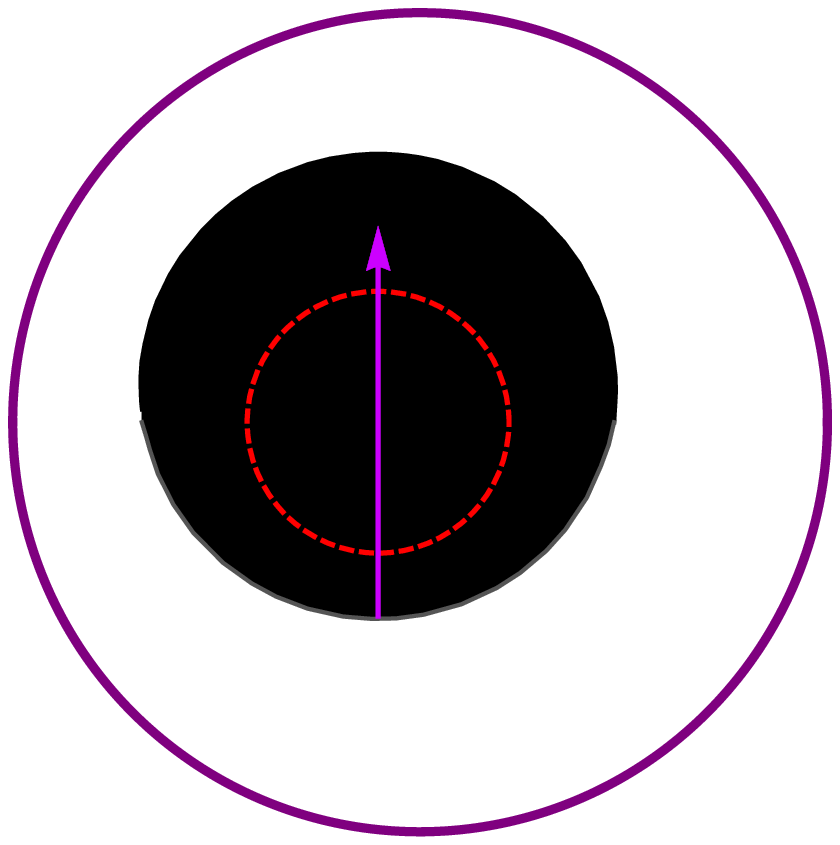}
	\hskip0.1cm
	\includegraphics[width=0.29\textwidth]{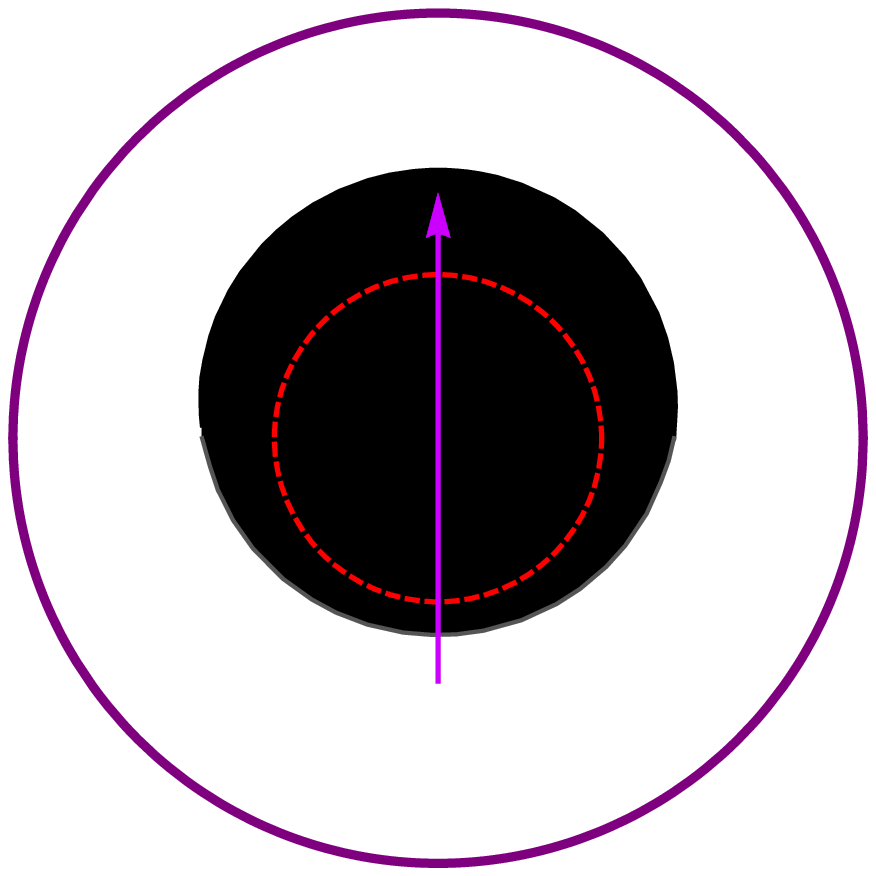}
	\caption{The dark silhouettes of the southern hemisphere of the event horizon (black region) in the case of supermassive black hole M87* ($\theta_0=17^\circ$), projected inside an outline of the classical black hole shadow (closed {\color{purple}purple} curves) in the cases of black hole spin, respectively, $a = 0.9982$ (left), $0.65$ (middle) and $0$ (right).}
	\label{fig10}      
\end{figure}

In this paper we use the (motivated by astrophysics) model of the black hole illuminated (highlighted) by a thin accretion disk in the black equatorial plane. In this model the outline (contour) of the dark event horizon silhouette (or, simply, the black hole silhouette) is defined by the highly red-shifted photons, emitted near the black hole event horizon by the thin accreting disk and registered by a distant observer. We calculate numerically the form of the black hole silhouette, which does not depend on the emission pattern of the thin accretion disk but completely determined by the black hole gravitational field. It appears that the form of the dark event horizon silhouette in this model is in good agreement with the form of the dark spot on the image of the supermassive black hole M87* obtained by the EHT collaboration.

Figure~\ref{fig7} from \cite{doknaz20} shows a visible dark silhouette of the northern hemisphere of the black hole event horizon illuminated by a thin accretion disk in the equatorial plane of the black hole with the spin $a=0.9982$, corresponding to the orientation of the supermassive black hole SgrA*. It is shown the typical photon trajectory (multicolored $3D$ curve), with parameters $\lambda=0.063$ and $q=0.121$, emitted by the hot accreting matter in the black hole equatorial plane at the radius $r=1.01\,r_{\rm h}$ and reaching a distant observer near the external boundary (contour) of the dark silhouette of the northern hemisphere of the event horizon globe. 

It is necessary to note that the motion of the accreting matter in the region inside the photon spheres and adjoining the event horizon is non-stationary. The accreting matter in this region is falling into the black hole along the spiraling down trajectories (see more details in \cite{doknazsm19,dokuch19,doknaz20}).

Figure~\ref{fig8} shows the possible forms of the dark event horizon silhouette (event horizon image) of the supermassive black hole SgrA* for three values of the black hole spin $a$. Note, that in the case of rotation axis orientation of the black hole SgrA* relative to a distant observer, the contour of the northern hemisphere of the lensed event horizon globe (black region) projected inside the boundary of classical  black hole shadow (closed purple region).

\begin{figure}
	\centering
	\includegraphics[width=0.6\textwidth,origin=c,angle=0]{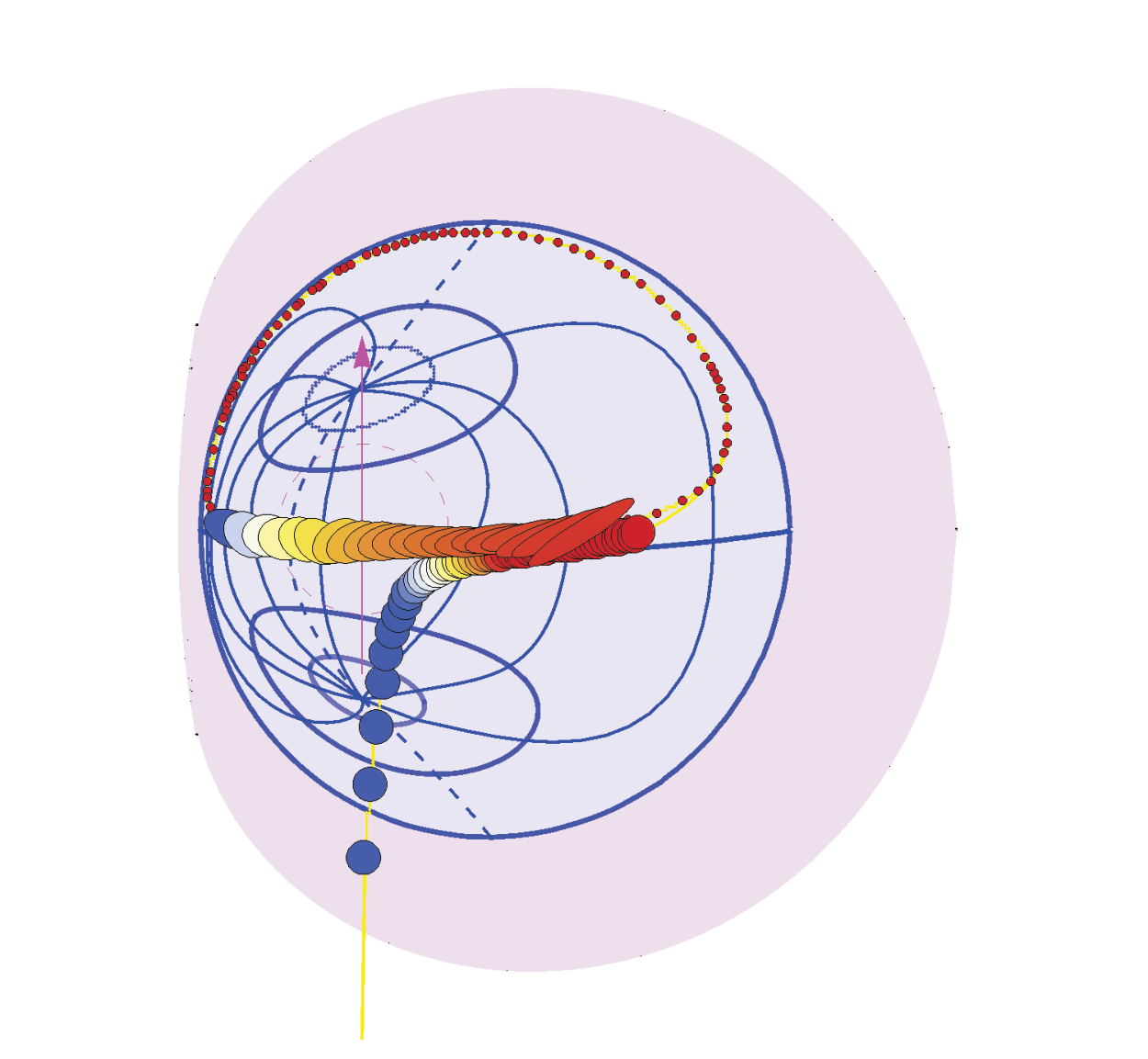}
	\caption{Numerical modeling of the gravitational lensing of the compact star, falling into the fast rotating black hole SgrA* ($a=0.9982$) and observed in discrete time intervals by a distant static observer, placed a little bit above the equatorial plane. The falling star has a zero azimuth angular momentum and moves in the black hole equatorial plane. An yellow curve is a viewed  trajectory of this star. The images of this star are projected on the celestial sphere inside the classical black hole shadow (a big closed light {\color{purple}purple} region), when this star is approaching the black hole event horizon, and then start to multiply winding up around the black hole very near to the black hole equatorial parallel $\theta=\pi/2$ on the lensed event horizon globe. It shown the first circle of this multiple winding. A brightness of the lensed star is exponentially  faded in time during successive windings. The closed {\color{blue}blue} curves are meridians and parallels on the reconstructed image of the lensed event horizon globe.}
	\label{fig11}
\end{figure}

A space orientation of the supermassive black hole M87* and its equatorial accretion disk relative to a distant observer at the Earth (or at the near-Earth space orbit) is shown in Figure~\ref{fig9}. The dashed magenta circle in this Figure and in all other similar ones corresponds to the black hole event horizon image in the imaginary Euclidean space. There are also shown two $3D$ photon trajectories, starting from the different points of the circle with a radius $r=1.01\,r_{\rm h}$ at the thin accretion disk (light green oval) in the equatorial plane of rotating black hole with the spin $a=0.9982$ and reaching a distant observer near the outer contour of the event horizon silhouette (dark gray region). Parameters of these two photon trajectories are $\lambda_1=-0.047$ and $q_1=2.19$ and, respectively, $\lambda_2=-0.029$ and $q_2=1.52$. 

The corresponding forms of the dark silhouette of the supermassive black holes SgrA* and M87*, highlighted by thin accretion disks, are shown, respectively in Figures~\ref{fig8} and \ref{fig10} for three different values of spin $a$. 

Note, that the dark event horizon silhouettes, similar to ones in Figures~\ref{fig7}--\ref{fig10}, were reproduced during many years in numerical modeling of accretion disks with the inner edge at the black hole event horizon (see, e.\,g., \cite{Luminet79,Bromley97,Fanton97,Fukue03,Fukue03b,Dexter09b,Ru-SenLu16,Luminet19,Shiokawa19,doknaz19,Gucht19,Kawashima19,White20,doknaz20,Lin20}). 

Figure~\ref{fig11} demonstrates a numerical model for the gravitational lensing of  compact star, falling into the fast rotating black hole SgrA* ($a=0.9982$) and observed in discrete time intervals by a distant static observer, placed a little bit above the equatorial plane. Falling star has a zero azimuth angular momentum and moves in the black hole equatorial plane. Images of this star are projected on the celestial sphere inside the classical black hole shadow (a big closed light {\color{purple}purple} region), when this star is approaching the black hole event horizon, and then are multiply winding up very near to the black hole equatorial parallel $\theta=\pi/2$ on the lensed event horizon globe. It is shown the first circle of this multiple winding. The brightness of the lensed star is exponentially  faded in time during successive windings (see animation in \cite{doknaz18c}). The closed {\color{blue}blue} curves are meridians and parallels on the reconstructed image of the lensed event horizon globe (for details see \cite{doknazsm19,dokuch19,doknaz20}).

\begin{figure}
	\centering
	\includegraphics[width=0.32\textwidth]{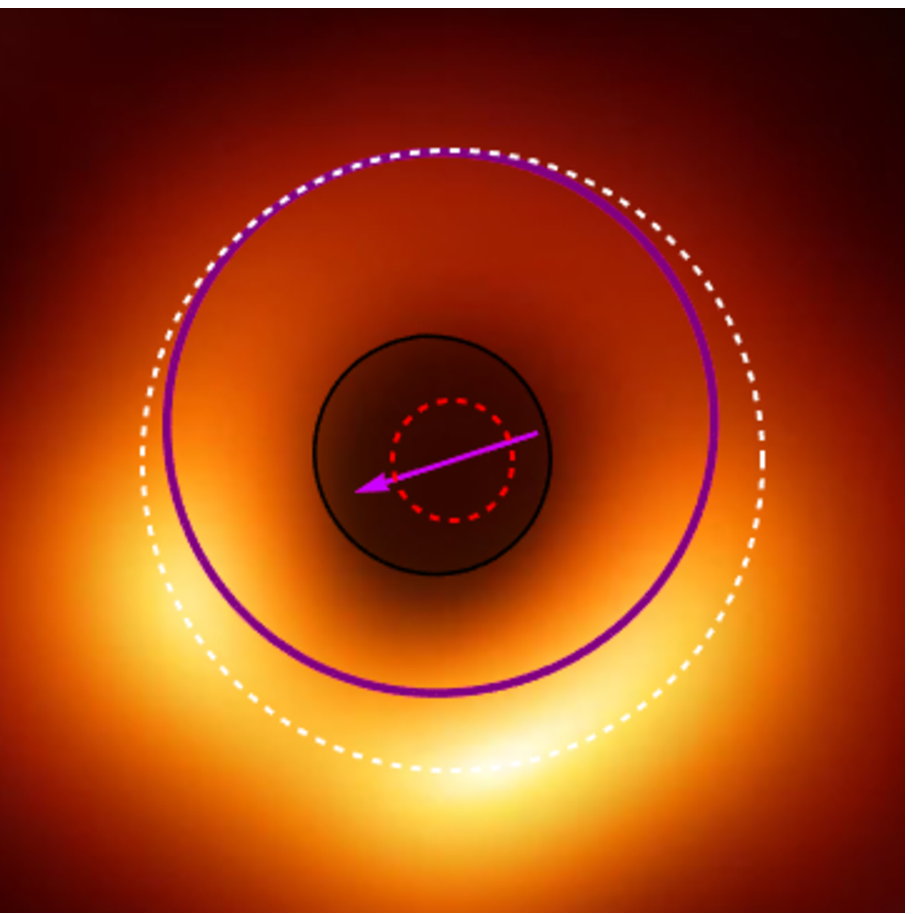}
	\hskip0.2 cm
	\includegraphics[width=0.32\textwidth]{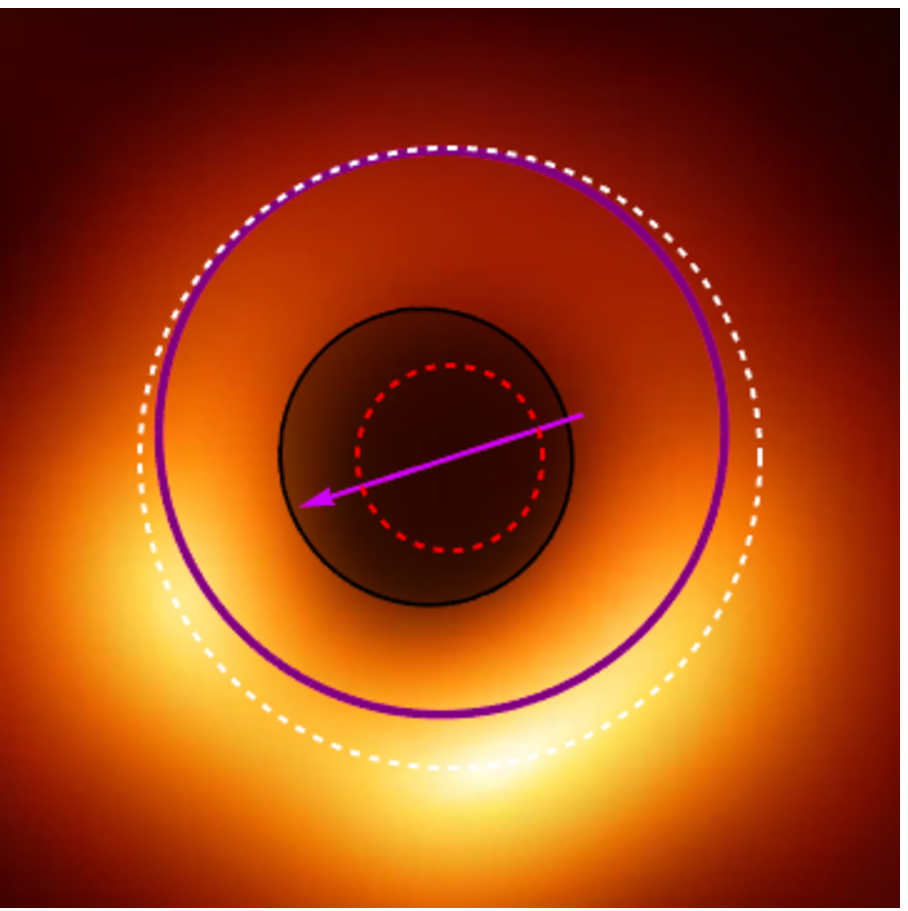}
	\hskip0.2 cm
	\includegraphics[width=0.32\textwidth]{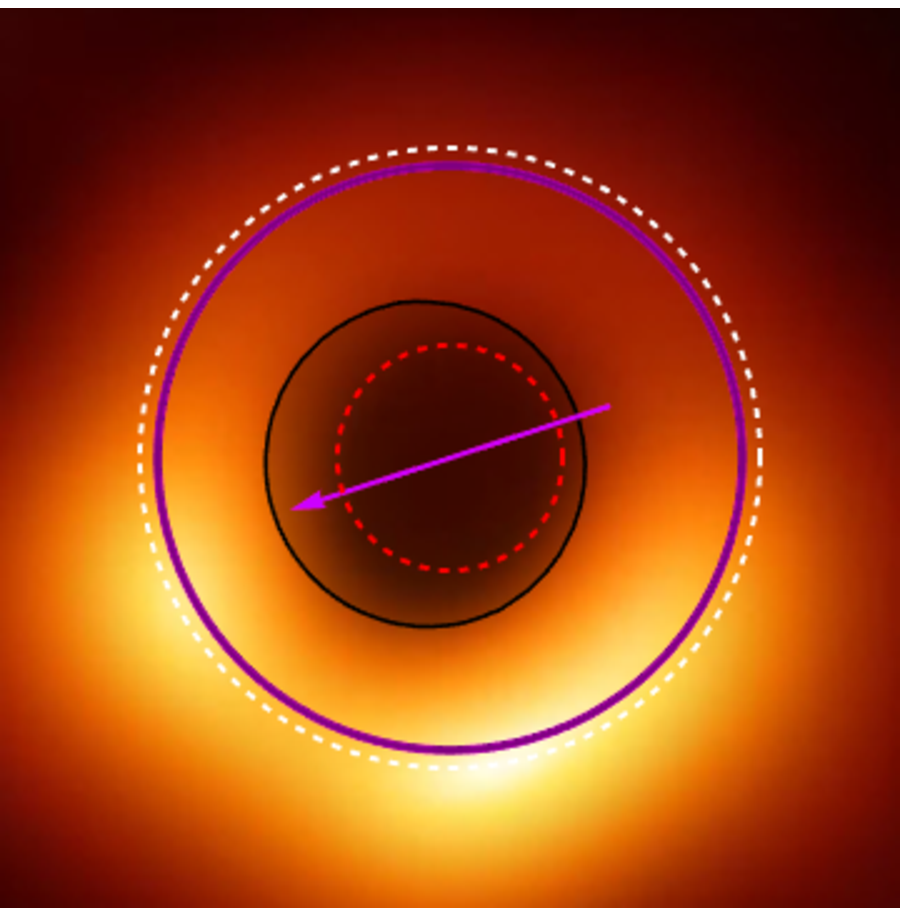}
	\caption{Superposition of the M87* image, obtained by the EHT, with both the contours af classical black hole shadows ({\color{purple}purple} closed curves) and the event horizon silhouettes (dark regions) for different values of the black hole spin, $a=0.998$ (left panel), $a=0.75$ (central panel) and $a=0$ (right panel). {\color{magenta}Magenta} arrows --- a black hole rotation axis. Small dashed {\color{red}ring} --- a black hole event horizon projection on the celestial sphere in the imaginary Euclidean space (in the absence of gravity). The modeled dark spot on the right panel in the $a=0$ case produces an excessively large dark spot in comparison with one on the EHT image. At the same time, a size of the dark spot on the EHT image agrees with a corresponding size of the dark event horizon silhouette in a thin accretion disk model in the case of either the high (left panel) or moderate (central panel) value of the black hole spin, $a\gtrsim0.75$.}
	\label{fig12}
\end{figure}

Finally, Figure~\ref{fig12} shows the superposition of the image of M87*, obtained by the EHT with both the contours of classical black hole shadows ({\color{purple}purple} closed curves) and the event horizon silhouettes (dark regions) for different values of the black hole spin, $a=0.998$ (left panel), $a=0.75$ (central panel) and $a=0$ (right panel). A white dashed circle with a radius $21\,\mu$as corresponds to $\approx5.5MG/c^2$. This circle was used by the EHT collaboration to model a bright crescent for the reconstruction of the M87* image \cite{EHT1}. The angular size of the M87* gravitational radius $MG/c^2$ corresponds to $\theta_{\rm }=3.8\,\mu$as. The modeled dark spot on the right panel in the  $a=0$ case produces an excessively large dark spot in comparison with one on the EHT image. We note that a size of the dark spot on the EHT image agrees with a corresponding size of the dark event horizon silhouette in a thin accretion disk model in the case of either the high or moderate value of the black hole spin, $a\gtrsim0.75$. Meantime, the EHT image details are blurred due to the observation resolution, and the direct identification of the edge of the bright crescent with the shadow or the silhouette can easily be misleading.

In the Figure~\ref{fig12} a position angle of the large-scale jet PA$=288^\circ$ and the viewing angle between the jet axis and line-of-sight $\theta_0=17^\circ$ according to \cite{Walker18,Nalewajko20}. We are especially grateful the authors of paper \cite{Nalewajko20} for pointing out the wrong angle of the large-scale jet from M87* PA$=215^\circ$, which we used for superposition of the EHT image of supermassive black hole M87* with the simulated event horizon silhouettes \cite{doknaz19b}. All model images, projected on the celestial sphere, must be rotated counter-clockwise to the angle $73^\circ$ to correct this error in \cite{doknaz19b}. Additionally, our claim in \cite{doknaz19b} on the preferable value of spin $a=0.75$ for M87* is  dubious. It seems that the dark silhouette of the event horizon on the image of M87* is either heavily shaded or accretion disk around this black hole is not thin.

\section{Discussions}

The classical black hole shadow is visible if the emission of luminous matter outside the photonic spheres dominates (e.\,g., if there is a distant luminous background of extended hot gas clouds or luminous stars far beyond the black hole). On the contrary, the much smaller event horizon silhouette is visible if the emission of luminous matter within the photonic spheres dominates (e.\,g., if there is a highly luminous accreting matter in the vicinity of the event horizon). 

The classical black hole shadow is complicated to observe with the present state of art either due to the low luminosity of the distant background far behind the black hole or due to the extremely high accretion activity of the black hole, which  completely dilute the emission from distant background. The basic  necessary requirement for observation of the black hole silhouette is, of course, the stability of the basic image structure among different days of observations according to \cite{EHT1}.

Photons, emitted near the event horizon by the luminous matter falling into the black hole, undergo extremely high red-shift by reaching a distant observer. Therefore, the registration accuracy of the event horizon silhouette strongly depends on the angular resolution and sensitivity of the used telescope.

A size of the dark spot on the EHT image agrees with a corresponding size of the dark event horizon silhouette in a thin accretion disk model (see Figure~\ref{fig12})) in the case of either the high or moderate value of the black hole spin, $a\gtrsim0.75$. The form of the dark event horizon silhouette does not depend on the emission pattern of the thin accretion disk but completely determined by the Kerr black hole gravitational field.

\begin{acknowledgements}
	We are grateful to E.O. Babichev, V.A. Berezin, Yu. N. Eroshenko and A.L. Smirnov for stimulating discussions. This study was supported in part by the Russian Foundation for Basic Research project 18-52-15001-NCNIa. 
\end{acknowledgements}


\begin{thebibliography}{999}
	
\bibitem{EHT1} The EHT Collaboration; Akiyama, K.; et al. The Event Horizon Telescope Collaboration. First M87 Event Horizon Telescope Results. I. The Shadow of the Supermassive Black Hole.  {\em Astrophys. J.} {\bf 2019},  875, L1, 17pp. 

\bibitem{EHT2} The EHT Collaboration; Akiyama, K.; et al. First M87 Event Horizon Telescope Results. II. Array and Instrumentation. {\em Astrophys. J.} {\bf 2019}, 875, L2, 28pp.

\bibitem{EHT3} The EHT Collaboration; Akiyama, K.; et al. First M87 Event Horizon Telescope Results. III. Data Processing and Calibration. {\em Astrophys. J.} {\bf 2019}, 875, L3, 32pp. 

\bibitem{EHT4} The EHT Collaboration; Akiyama, K.; et al.  First M87 Event Horizon Telescope Results. IV. Imaging the Central Supermassive Black Hole. {\em Astrophys. J.} {\bf 2019}, 875, L4, 52pp. 

\bibitem{EHT5} The EHT Collaboration; Akiyama, K.; et al. First M87 Event Horizon Telescope Results. V. Physical Origin of the Asymmetric Ring. {\em Astrophys. J.} {\bf 2019}, 875, L5, 31pp. 

\bibitem{EHT6} The EHT Collaboration; Akiyama, K.; et al. First M87 Event Horizon Telescope Results. VI. The Shadow and Mass of the Central Black Hole.  {\em Astrophys. J.} {\bf 2019}, 875, L6, 44pp.

\bibitem{Kerr} Kerr, R. P. Gravitational Field of a Spinning Mass as an Example of Algebraically Special Metrics.  {\em Phys. Rev. Lett.} {\bf 1963}, 11, 237-238. 

\bibitem{Fish16} Fish, V. L.; et al. Observing—and Imaging—Active Galactic Nuclei with the Event Horizon Telescope. {\em Galaxies} {\bf 2016},  4(4), 54.

\bibitem{Lacroix13} Lacroix, T.; Silk, J. Constraining the distribution of dark matter at the Galactic centre using the high-resolution Event Horizon Telescope. {\em Astron. Astrophys.} {\bf 2013},   554, A36.

\bibitem{Kamruddin} Kamruddin, A. B.; Dexter, J. A geometric crescent model for black hole images. {\em Mon. Not. R. Astron. Soc.} {\bf 2013 }, 434, 765-771.

\bibitem{Johannsen16} Johannsen, T.; et al. Testing General Relativity with the Shadow Size of SgrA*. {\em Phys. Rev. Lett.} {\bf 2016}, 116, 031101.

\bibitem{Johannsen16b} Johannsen, T.; et al. Testing General Relativity with Accretion-Flow Imaging of SgrA*. {\em Phys. Rev. Lett.} {\bf 2016}, 117, 091101.

\bibitem{Broderick16} Broderick, A. E.; et al. Modeling Seven Years of Event Horizon Telescope Observations with Radiatively Inefficient Accretion Flow Models. {\em Astrophys. J.} {\bf 2016}, 820, 137, 16pp.

\bibitem{Chael16}  Chael, A. A.; et al. High Resolution Linear Polarimetric Imaging for the Event Horizon Telescope. {\em Astrophys. J.} {\bf 2016}, 829, 11.

\bibitem{Kim16} Kim, J.; et al. Bayesian techniques for comparing time-dependent GRMHD simulations to variable Event Horizon Telescope observations.  {\em Astrophys. J.} {\bf 2016}, 832, 156.

\bibitem{Roelofs17} Roelofs, F.; et al. Quantifying Intrinsic Variability of Sagittarius A* Using Closure Phase Measurements of the Event Horizon Telescope. {\em Astrophys. J.} {\bf 2017}, 847, 55.

\bibitem{Doeleman17} Doeleman, S. S. Seeing the unseeable.  {\em Nat. Astron.} {\bf 2017}, 1, 646.

\bibitem{Gheetal08} Ghez, A. M.; et al. Measuring Distance and Properties of the Milky Way's Central Supermassive Black Hole with Stellar Orbits.  {\em Astrophys. J.} {\bf 2008}, 689, 1044-1062.

\bibitem{Gillessen09} Gillessen, S.; et al.  Monitoring stellar orbits around the Massive Black Hole in the Galactic Center. {\em Astrophys. J.} {\bf 2009},  692,  1075-1109.

\bibitem{Giletal09-2} Gillessen, S.; et al. The orbit of the star S2 around SGR A* from very large telescope and Keck data. {\em Astrophys. J.} {\bf 2009},  707, L114-L117.

\bibitem{Meyer12} Meyer, L.; et al. The Shortest-Known-Period Star Orbiting Our Galaxy’s Supermassive Black Hole. {\em Science} {\bf 2012}, 338, 84-87.

\bibitem{Johannsen12} Johannsen, T.; et al. Masses of nearby Supermassive Black Holes with Very Long Baseline Interferometry.  {\em Astrophys. J.} {\bf 2012}, 758, 30-37.

\bibitem{Baade46} Baade, W. A Search For the Nucleus of Our Galaxy. {\em Publ. Astron. Soc. Pacific} {\bf 1946}, 58, 249-252.

\bibitem{Becklin68} Becklin, E. E.; Neugebauer, G. Infrared Observations of the Galactic Center.  {\em Astrophys. J.} {\bf 1968}, 151, 145.

\bibitem{Dokuchaev77} Dokuchaev, V. I.; Ozernoi, L. M. Tidal disruption of stars and the evolution of a massive black hole under the conditions of the galactic center. {\em Soviet Astron. Lett.} {\bf 1977}, 3, 209-211; Pis'ma Astron. Zh. {\bf 1977}, 3, 91-395.

\bibitem{Dokuchaev89} Dokuchaev, V. I. The Evolution of a Massive Black-Hole in the Nucleus of a Normal Galaxy. {\em Sov. Astron. Lett.} {\bf 1989}, 15, 167-175; Pis'ma Astron. Zh. {\bf 1989}, 15, 387-395.

\bibitem{Allen90} Allen, D. A.; Hyland, A. R.; Hillier, D. J. The source of luminosity at the Galactic Centre.  {\em Mon. Not. R. Astron. Soc.} {\bf 1990}, 244, 706-713.

\bibitem{Dokuchaev91} Dokuchaev, V. I. Joint evolution of a galactic nucleus and central massive black hole. {\em Mon. Not. R. Astron. Soc.} {\bf 1991}, 251, 564-574.

\bibitem{Dokuchaev91b} Dokuchaev, V. I. Birth and life of massive black holes. {\em Sov. Phys. Usp.} {\bf 1991}, 34, 447–470; Usp. Fiz. Nauk {\bf 1991}, 161, 1–52.

\bibitem{Manko92} Manko, V. S.; Novikov, I. D. Generalizations of the Kerr and Kerr-Newman metrics possessing an arbitrary set of mass-multipole moments.   {\em Class. Quantum Grav.} {\bf 1992}, 9, 2477-2487.

\bibitem{Lo93} Lo, K. Y.; et al.  High-resolution VLBA imaging of the radio source SgrA* at the Galactic Centre. {\em Nature} {\bf 1993}, 362, 38–40.

\bibitem{Backer93} Backer, D. C.; et al. Upper limit of 3.3 astronomical units to the diameter of the galactic center radio source SgrA*. {\em Science} {\bf 1993}, 262, 1414-1416.

\bibitem{Eckart96} Eckart, A.; Genzel, R. Observations of stellar proper motions near the Galactic Centre. {\em Nature} {\bf 1996}, 383, 415–417.

\bibitem{Haller96} Haller, J. W.; et al. Stellar Kinematics and the Black Hole in the Galactic Center: Erratum.  {\em Astrophys. J.} {\bf 1996}, 468, 955.

\bibitem{Ghez98} Ghez, A. M.; et al. High Proper-Motion Stars in the Vicinity of Sagittarius A*: Evidence for a Supermassive Black Hole at the Center of Our Galaxy. {\em Astrophys. J.} {\bf 1998}, 509, 678-686. 

\bibitem{Backer99} Backer, D. C.; Sramek, R. A. Proper Motion of the Compact, Nonthermal Radio Source in the Galactic Center, Sagittarius A*. {\em Astrophys. J.} {\bf 1999}, 524, 805-815.

\bibitem{Reid99} Reid, M. J.; et al.  The Proper Motion of Sagittarius A*. I. First VLBA Results. {\em Astrophys. J.} {\bf 1999}, 524, 816-823.

\bibitem{Baganoff99} Baganoff, F. K.; et al. Chandra Imaging of SgrA* and the Galactic Center. {\em Bull. Am. Astron. Soc.} {\bf 1999}, 31, 1463.

\bibitem{Falcke00b} Falcke, H.; Markoff, S.  The jet model for Sgr A*: Radio and X-ray spectrum. {\em Astron. Astrophys.} {\bf 2000}, 362, 113-118.

\bibitem{NovikovFrolov01} Novikov, I. D.;  Frolov, V. P. Black holes in the Universe.  {\em Phys. Usp.} {\bf 2001}, 44, 291-305; Usp. Fiz. Nauk {\bf 2001}, 171, 307-324.  

\bibitem{Baganoff01} Baganoff, F. K.; et al. Rapid X-ray flaring from the direction of the supermassive black hole at the Galactic Centre. {\em Nature} {\bf 2001}, 413, 45-48. 

\bibitem{Hornstein02} Hornstein, S. D.; et al. Limits on the Short-Term Variability of Sagittarius A* in the Near-Infrared.  {\em Astrophys. J. Lett.} {\bf 2002}, 577, L9-L13.

\bibitem{Genzel03} Genzel, R.; et al. Near-infrared flares from accreting gas around the supermassive black hole at the Galactic Centre. {\em Nature} {\bf 2003}, 425, 934-937.

\bibitem{Aschenbach04} Aschenbach, B.; et al. X-ray flares reveal mass and angular momentum of the Galactic Center black hole. {\em Astron. Astrophys.} {\bf 2004}, 417, 71-78.

\bibitem{YusefZadeh06} Yusef-Zadeh, F.; et al. Flaring Activity of Sagittarius A* at 43 and 22 GHz: Evidence for Expanding Hot Plasma. {\em Astrophys. J.} {\bf 2006}, 650, 189-194.

\bibitem{Marrone08} Marrone, D. P.; et al. An X-Ray, Infrared, and Submillimeter Flare of Sagittarius A*. {\em Astrophys. J.} {\bf 2008}, 682, 373-383. 

\bibitem{Ghez08} Ghez, A. M.; et al. Measuring Distance and Properties of the Milky Way's Central Supermassive Black Hole with Stellar Orbits. {\em Astrophys. J.} {\bf 2008}, 689, 1044-1062.

\bibitem{Doeleman08} Doeleman, S. S.; et al. Event-horizon-scale structure in the supermassive black hole candidate at the Galactic Centre. {\em Nature} {\bf 2008}, 455, 78–80.

\bibitem{Doeleman09} Doeleman, S. S.; et al. Detecting Flaring Structures in Sagittarius A* with High-Frequency VLBI. {\em Astrophys. J.} {\bf 2009}, 695, 59-74.

\bibitem{DoddsEden09} Dodds-Eden, K.; et al. Evidence for X-ray synchrotron emission from simultaneous mid-IR to X-ray observations of a strong Sgr A* flare. {\em Astrophys. J.} {\bf 2009}, 698, 676-692.

\bibitem{Broderick09} Broderick, A. E.; Loeb, A.; Narayan, R. The Event Horizon of Sagittarius A*. {\em Astrophys. J.} {\bf 2009}, 701, 1357-1366.

\bibitem{Sabha10} Sabha, N.; et al. The extreme luminosity states of Sagittarius A*. {\em Astron. Astrophys. J.} {\bf 2010}, 512, A2. 

\bibitem{Dexter10} Dexter, J.; et al. The Submillimeter Bump in Sgr A* from Relativistic MHD Simulations.  {\em Astrophys. J.} {\bf 2010}, 717, 1092-1104.

\bibitem{Paolis11} De Paolis,  F.; et al. Estimating the parameters of the Sgr A* black hole. {\em Gen. Relativ. Gravit.} {\bf 2011}, 43, 977–988. 

\bibitem{Broderick11} Broderick, A. E.; Loeb, A.; Reid, M. J. Localizing Sagittarius A* and M87 on Microarcsecond Scales with Millimeter Very Long Baseline Interferometry. {\em Astrophys. J.} {\bf 2011}, 735, 57-75.

\bibitem{Neilsen13} Neilsen, J.; et al. A Chandra/HETGS Census of X-Ray Variability from Sgr A* during 2012. {\em Astrophys. J.} {\bf 2013}, 774, 42-56. 

\bibitem{Zakharov13} Borka, D.; et al.  Constraining the range of Yukawa gravity interaction from S2 star orbits. {\em JCAP} {\bf 2013}, 11, 050. 

\bibitem{Fish14} Fish, V. L.; et al. Imaging an Event Horizon: Mitigation of Scattering Toward Sagittarius A*. {\em Astrophys. J.} {\bf 2014}, 795, 134-141.  

\bibitem{Gwinn14} Gwinn, C. R.; et al. Discovery of Substructure in the Scatter-Broadened Image of Sgr A*. {\em Astrophys. J. Lett.} {\bf 2014}, 794, L14-L19. 

\bibitem{Johnson14} Johnson, M. D.; et al. Relative Astrometry of Compact Flaring Structures in Sgr A* with Polarimetric Very Long Baseline Interferometry. {\em Astrophys. J. Lett.} {\bf 2014}, 794, 150-158. 

\bibitem{Dokuch14} Dokuchaev, V. I. Spin and mass of the nearest supermassive black hole. {\em Gen. Relativ. Gravit.} {\bf 2014}, 46, 1832-1845.

\bibitem{Moscibrodzka14} Mo\'scibrodzka, M.; et al. Observational appearance of inefficient accretion flows and jets in 3D GRMHD simulations: Application to Sagittarius A*.  {\em Astron. Astrophys.} {\bf 2014}, 570, A7.

\bibitem{Bower15} Bower, G. C.; et al. Radio and Millimeter Monitoring of Sgr A*: Spectrum, Variability, and Constraints on the G2 Encounter. {\em Astrophys. J.} {\bf 2015}, 802, 69-83.

\bibitem{Johnson15} Johnson, M. D.; et al. Resolved magnetic-field structure and variability near the event horizon of Sagittarius A*. {\em Science} {\bf 2015}, 350, 1242-1245.

\bibitem{Chatzopoulos15} Chatzopoulos, S.; et al. The old nuclear star cluster in the Milky Way: dynamics, mass, statistical parallax, and black hole mass. {\em Mon. Not. R. Astron. Soc.} {\bf 2015}, 447,  948–968.

\bibitem{FizLab} Dokuchaev, V. I.; Eroshenko, Yu. N.  Physical laboratory at the center of the Galaxy. {\em Phys. Usp.} {\bf 2015}, 58, 772–784;  Usp. Fiz. Nauk {\bf 2015}, 185, 829–843. 

\bibitem{Rauch16}  Rauch, C.; et al.  Wisps in the Galactic center: Near-infrared triggered observations of the radio source Sgr A* at 43 GHz. {\em Astron. Astrophys. J.} {\bf 2016},  587, A37. 

\bibitem{Zakharov16} Zakharov, A. F.; et al. Constraining the range of Yukawa gravity interaction from S2 star orbits II: bounds on graviton mass. {\em JCAP} {\bf 2016}, 05, 045.

\bibitem{Becerril16} Becerril, R.; Valdez-Alvarado, S.; Nucamendi, U. Obtaining mass parameters of compact objects from redshifts and blueshifts emitted by geodesic particles around them. {\em Phys. Rev. D} {\bf 2016}, 94,  124024. 

\bibitem{Giddings16} Giddings, S. B.; Psaltis, D. Event Horizon Telescope observations as probes for quantum structure of astrophysical black holes.  {\em Phys. Rev. D} {\bf 2018}, 97, 084035.

\bibitem{Johannsen16c} Johannsen, T. Sgr A* and General Relativity. {\em Class. Quantum Grav.} {\bf 2016}, 33,  113001.

\bibitem{OrtizLeon16} Parsa Ortiz-Le\'on, G. N.; et al. The Intrinsic Shape of Sagittarius A* at 3.5-mm Wavelength. {\em Astrophys. J.} {\bf 2016}, 824, 40-50. 

\bibitem{Parsa17} Parsa, M.; et al.  Investigating the Relativistic Motion of the Stars Near the Supermassive Black Hole in the Galactic Center. {\em Astrophys. J.} {\bf 2017}, 845, 22-41.

\bibitem{Capellupo17} Capellupo, D. M.; et al. Simultaneous Monitoring of X-Ray and Radio Variability in Sagittarius A*. {\em Astrophys. J.} {\bf 2017}, 845, 35. 

\bibitem{Shiokawa17} Shiokawa, H.; Gammie, C. F.; Doeleman, S.  Time Domain Filtering of Resolved Images of Sgr A*. {\em Astrophys. J.} {\bf 2017}, 846,  29-41. 

\bibitem{Johnson17} Johnson, M. D.; et al. Dynamical Imaging with Interferometry. {\em Astrophys. J.} {\bf 2017}, 850, 172-187.

\bibitem{Eckart17} Eckart, A.; et al. The Milky Way's Supermassive Black Hole: How Good a Case Is It? {\em Found. Phys.} {\bf 2017}, 47, 553-624.

\bibitem{Abdujabbarov17} Abdujabbarov, A.; et al.  Shadow of the rotating black hole with quintessential energy in the presence of plasma. {\em Int. J. Mod. Phys. D} {\bf 2017}, 26, 1750051.

\bibitem{Ponti17} Ponti, G.; et al. A powerful flare from Sgr A* confirms the synchrotron nature of the X-ray emission. {\em Mon. Not. R. Astron. Soc.} {\bf 2017}, 468, 2447–2468. 

\bibitem{Zajacek18} Zaja\v{c}ek, M.; et al.  On the charge of the Galactic centre black hole. {\em Mon. Not. R. Astron. Soc.} {\bf 2018}, 480, 4408-4423. 

\bibitem{Abuter18} GRAVITY Collaboration; Abuter, R.; et al. Detection of orbital motions near the last stable circular orbit of the massive black hole SgrA*. {\em Astron. Astrophys.} {\bf 2018}, 618, L10.

\bibitem{Zakharov18a} Zakharov, A. F. The black hole at the Galactic Center: Observations and models. {\em Int. J. Mod. Phys. D} {\bf 2018}, 27, 1841009.

\bibitem{Zakharov18b} Zakharov, A. F. Constraints on tidal charge of the supermassive black hole at the Galactic Center with trajectories of bright stars. {\em Eur. Phys. J. C} {\bf 2018}, 78, 689.

\bibitem{Zakharov18c} Zakharov, A. F.  Constraints on alternative theories of gravity with observations of the Galactic Center. {\em EPJ Web Conf.} {\bf 2018}, 191, 01010.

\bibitem{Zhu19} Zhu, Z.; et al. Testing General Relativity with the Black Hole Shadow Size and Asymmetry of Sagittarius A*: Limitations from Interstellar Scattering. {\em Astrophys. J.} {\bf 2019}, 870, 6.

\bibitem{Izmailov19} Izmailov, R. N.; et al. Can massless wormholes mimic a Schwarzschild black hole in the strong field lensing? {\em Eur. Phys. J. Plus} {\bf 2019}, 134, 384. 

\bibitem{Zakharov19} Zakharov, A. F. Tests of gravity theories with Galactic Center observations. {\em Int. J. Mod. Phys. D} {\bf 2019}, 28,  1941003.

\bibitem{TuanDo19} Do, T.; et al. Envisioning the next decade of Galactic Center science: a laboratory for the study of the physics and astrophysics of supermassive black holes. {\em arXiv}  {\bf 2019}, {arXiv:1903.05293}.

\bibitem{Do19} Do, T.; et al. Unprecedented variability of SgrA* in NIR. {\em arXiv}  {\bf 2019}, {arXiv:1908.01777}.

\bibitem{Giddings19} Giddings, S. B. Searching for Quantum Black Hole Structure with the Event Horizon Telescope. {\em Universe} {\bf 2019},  5(9), 201. 

\bibitem{Dai19} Dai, D.-C.; Stojkovic, D. Observing a wormhole.  {\em Phys. Rev. D} {\bf 2019}, 100,  083513. 

\bibitem{Moriyama19} Moriyama, K.; et al. Black Hole Spin Measurement Based on Time-domain VLBI Observations of Infalling Gas Clouds.  {\em Astrophys. J.} {\bf 2019},  887, 227.

\bibitem{Ho08} Ho, L. C. Nuclear Activity in Nearby Galaxies. {\em Annu. Rev. Astron. Astrophys.} {\bf 2008}, 46, 475-539.

\bibitem{Gebhardt09} Gebhardt, K.; Thomas, J. The Black Hole Mass, Stellar Mass-to-Light Ratio, and Dark Halo in M87. {\em Astrophys. J.} {\bf 2009}, 700, 1690–1701.

\bibitem{Gebhardt11} Gebhardt, K.; et al. The Black-Hole Mass in M87 from Gemini/NIFS Adaptive Optics Observations. {\em Astrophys. J.} {\bf 2011}, 729, 119-132. 

\bibitem{Walsh13} Walsh, J.; et al. The M87 Black Hole Mass from Gas-dynamical Models of Space Telescope Imaging Spectrograph Observations. {\em Astrophys. J.} {\bf 2013}, 770, 86-97.

\bibitem{Goddi17} Goddi, C.; et al. BlackHoleCam: fundamental physics of the Galactic center. {\em Int. J. Mod. Phys. D} {\bf 2017}, 26, 1730001. 

\bibitem{GRAVITY18} GRAVITY Collaboration; Abuter, R.; et al.  Detection of the gravitational redshift in the orbit of the star S2 near the Galactic centre massive black hole. {\em Astron. Astrophys.} {\bf 2018},  615, L15.

\bibitem{GRAVITY19} GRAVITY Collaboration; Amorim, A.; et al.  Test of the Einstein Equivalence Principle near the Galactic Center Supermassive Black Hole. {\em Phys. Rev. Lett.} {\bf 2019}, 122,  101102. 

\bibitem{Mielnik62} Mielnik, B.; Pleba\'nski, J. A Study of Geodesic Motion in the Field of Schwarzchild’s Solution. {\em Acta Phys. Pol.} {\bf 1962}, 21,  239–268.

\bibitem{Synge63} Synge, J. L. The Escape of Photons from Gravitationally Intense Stars. {\em Mon. Not. R. Astron. Soc.} {\bf 1966}, 131, 463–466.

\bibitem{Bardeen73} Bardeen, J. M. in {\em Black Holes}; DeWitt, C., DeWitt, B. S., Eds.; Gordon and Breach Science Publishers:  New York, United States, 1973; pp. 215–239.

\bibitem{Young76} Young, P. I. Capture of particles from plunge orbits by a black hole. {\em Phys. Rev. D} {\bf 1976}, 14,  3281. 

\bibitem{Chandra} Chandrasekhar, S. Chapter 7. {\em The Mathematical Theory of Black Holes}. (The International Series of Monograph on Physics, Vol. 69); Clarendon Press: Oxford, 1983.  

\bibitem{Falcke00} Falcke, H.; Melia, F.; Agol, E. Viewing the Shadow of the Black Hole at the Galactic Center. {\em Astrophys. J.} {\bf 2000}, 528, L13-L16. 

\bibitem{Takahashi04}  Takahashi, R. Shapes and Positions of Black Hole Shadows in Accretion Disks and Spin Parameters of Black Holes. {\em Astrophys. J.} {\bf 2004}, 611, 996-1004. 

\bibitem{Kardashev07} Kardashev, N. S.;  Novikov, I. D.;  Shatskiy, A. A. Astrophysics of Wormholes. {\em Int. J. Mod. Phys. D} {\bf 2007}, 16, 909-926. 

\bibitem{Falcke13} Falcke, H.; Markoff, S. Toward the event horizon—the supermassive black hole in the Galactic Center.  {\em Class. Quantum Grav.} {\bf 2013}, 30, 244003. 

\bibitem{Li14} Li, Z.; Bambi, C. Testing SgrA* with the spectrum of its accretion structure. {\em JCAP} {\bf 2014}, 01, 041. 

\bibitem{Inoue14} Inoue, M.; et al. Greenland telescope project: Direct confirmation of black hole with sub‐millimeter VLBI.  {\em Radio Sci.} {\bf 2014}, 49, 564-571. 

\bibitem{Cunha15} Cunha, P. V. P.; et al.  Shadows of Kerr Black Holes with Scalar Hair. {\em Phys. Rev. Lett.} {\bf 2015}, 115, 211102. 

\bibitem{Abdujabbarov15} Abdujabbarov, A. A.; Rezzolla, L.; Ahmedov, B. J.  A coordinate-independent characterization of a black hole shadow.  {\em Mon. Not. R. Astron. Soc.} {\bf 2015}, 454, 2423–2435. 

\bibitem{Younsi16} Younsi, Z.; et al.  A new method for shadow calculations: application to parameterised axisymmetric black holes. {\em Phys. Rev. D} {\bf 2016}, 94,  084025.

\bibitem{deVries00} de Vries, A. The apparent shape of a rotating charged black hole, closed photon orbits and the bifurcation set $A_4$. {\em Class. Quantum Grav.} {\bf 2000}, 17,  123.

\bibitem{Schnittman06} Schnittman, J. D.; Krolik, J. H.; Hawley, J. F. Light Curves from an MHD Simulation of a Black Hole Accretion Disk.   {\em Astrophys. J.} {\bf 2006}, 651, 1031-1048. 

\bibitem{Shatskiy08} Shatskiy, A. A.; Novikov, I. D.; Kardashev, N. S. A dynamic model of the wormhole and the Multiverse model. {\em Phys. Usp.} {\bf 2008}, 51, 457–464;  Usp. Fiz. Nauk {\bf 2008}, 178, 481–488. 

\bibitem{Bambi09} Bambi, C.; Freese, K. Apparent shape of super-spinning black holes. {\em Phys. Rev. D} {\bf 2009}, 79, 043002. 

\bibitem{Frolov09} Frolov, V. P.; Shapiro, I. L. Black holes in higher dimensional gravity theory with corrections quadratic in curvature. {\em Phys. Rev. D} {\bf 2009}, 80, 044034. 

\bibitem{Tamburini11} Tamburini, F.; et al. Twisting of light around rotating black holes. {\em Nat. Phys.} {\bf 2011}, 7, 195-197. 

\bibitem{Vincent11} Vincent, F. H.; et al.  GYOTO: a new general relativistic ray-tracing code. {\em Class. Quantum Grav.} {\bf 2011}, 28, 225011.  

\bibitem{Amarilla12} Amarilla, L.; Eiroa, E. F. Shadow of a rotating braneworld black hole.  {\em Phys. Rev. D} {\bf 2012}, 85,  064019. 

\bibitem{Johannsen13} Johannsen, T. Photon Rings around Kerr and Kerr-like Black Holes. {\em Astrophys. J} {\bf 2013}, 777, 170-182. 

\bibitem{Babichev13} Babichev, E. O.; Dokuchaev, V. I.; Eroshenko, Yu. N.  Black holes in the presence of dark energy. {\em Phys. Usp.} {\bf 2013}, 56, 1155–1175; Usp. Fiz. Nauk {\bf 2013}, 183, 1257–1280. 

\bibitem{Amarilla13} Amarilla, L.; Eiroa, E. F. Shadow of a Kaluza-Klein rotating dilaton black hole. {\em Phys. Rev. D} {\bf 2013}, 87, 044057.

\bibitem{Zakharov14} Zakharov, A. F.; et al. Constraints on RnRn gravity from precession of orbits of S2-like stars: a case of a bulk distribution of mass. {\em Adv. Space Res.} {\bf 2014}, 54, 1108-1112. 

\bibitem{DokEr15} Dokuchaev, V. I.; Eroshenko, Yu. N. Weighing of the dark matter at the center of the galaxy. {\em JETP Lett.} {\bf 2015}, 101, 777–782; Pis'ma Zh. Eksp. Teor. Fiz. {\bf 2015}, 101, 875–880. 

\bibitem{Wei15} Wei, S.-W.; et al. Shadow of noncommutative geometry inspired black hole. {\em JCAP} {\bf 2015}, 08,  004. 

\bibitem{Abd15} Abdolrahimi, S.; Mann, R. B.;  Tzounis, C.  Distorted local shadows.  {\em Phys. Rev. D} {\bf 2015}, 91,  084052. 

\bibitem{Nucamendi15} Herrera-Aguilar, A.; Nucamendi, U. Kerr black hole parameters in terms of the redshift/blueshift of photons emitted by geodesic particles. {\em Phys. Rev. D} {\bf 2015}, 92, 045024.

\bibitem{Nucamendi16} Becerril, R.; Valdez-Alvarado, S.; Nucamendi, U. Obtaining mass parameters of compact objects from redshifts and blueshifts emitted by geodesic particles around them. {\em Phys. Rev. D} {\bf 2016}, 94, 124024. 

\bibitem{Cunha16} Cunha, P. V. P.; et al. Shadows of Kerr black holes with and without scalar hair. {\em Int. J. Mod. Phys. D} {\bf 2016}, 25,  1641021.

\bibitem{Abdujabbarov16} Abdujabbarov, A. A.; et al. Shadow of rotating regular black holes. {\em Phys. Rev. D} {\bf 2016}, 93, 104004. 

\bibitem{CliffWill17a} Will, C. M.; Maitra, M. Relativistic orbits around spinning supermassive black holes: Secular evolution to 4.5 post-Newtonian order. {\em Phys. Rev. D} {\bf 2017}, 95,  064003. 

\bibitem{Cunha17} Cunha, P. V. P.; Herdeiro, C.; Radu, E. Fundamental photon orbits: Black hole shadows and spacetime instabilities. {\em Phys. Rev. D} {\bf 2017}, 96,  024039. 

\bibitem{CliffWill17b} Ferrer, F.; Medeiros da Rosa, A.; Will, C. M.   Dark matter spikes in the vicinity of Kerr black holes.  {\em Phys. Rev. D} {\bf 2017}, 96, 083014.

\bibitem{Amarilla17} Amarilla, L.; Eiroa, E. F. in {\em The Fourteenth Marcel Grossmann Meeting. Proc. of the MG14 Meeting 0n General Relativity, Univ. of Rome La Sapienza, Italy, 12-18 July 2015}; Bianchi, M.,  Jantzen, R. T.,  Ruffini, R., Eds.;  World Scientific: Singapore, 2017; p. 3543. %

\bibitem{Mureika17} Mureika, J. R.; Varieschi, G. U. Black hole shadows in fourth-order conformal Weyl gravity. {\em Can. J. Phys} {\bf 2017}, 95, 1299–1306.

\bibitem{Amir18} Amir, M.; Singh, B. P.; Ghosh, S. G. Shadows of rotating five-dimensional charged EMCS black holes. {\em Eur. Phys. J. C} {\bf 2018}, 78, 399.

\bibitem{Lan18} Lan, X. G.; Pu, J. Observing the contour profile of a Kerr–Sen black hole. {\em Mod. Phys. Lett. A} {\bf 2018}, 33, 1850099.

\bibitem{Wang18}  Wang, M.; Chen, S.; Jing, J. Chaotic shadow of a non-Kerr rotating compact object with quadrupole mass moment. {\em Phys. Rev. D} {\bf 2018}, 98, 104040.

\bibitem{Lamy18} Lamy, F.; et al. Imaging a non-singular rotating black hole at the center of the Galaxy. {\em Class. Quantum Grav.} {\bf 2018}, 35, 115009. 

\bibitem{Mizuno18} Mizuno, Y.; et al. The Current Ability to Test Theories of Gravity with Black Hole Shadows. {\em Nature Astronomy} {\bf 2018}, 2, 585. 

\bibitem{Repin18} Repin, S. V.; et al. Shadow of rotating black holes on a standard background screen. {\em arXiv} {\bf 2019}, {arXiv:1802.04667}. 

\bibitem{Hennigar18} Hennigar, R. A.; Poshteh, M. B. J.; Mann, R. B. Shadows, Signals, and Stability in Einsteinian Cubic Gravity.  {\em Phys. Rev. D} {\bf 2018}, 97, 064041. 

\bibitem{Wei19} Wei, S. -W.; et al. Curvature radius and Kerr black hole shadow. {\em JCAP} {\bf 2019}, 08,  030. 

\bibitem{Blackburn19} Blackburn, L.; et al. {\em Astro2020 APC White Paper} {\bf 2020}.  

\bibitem{Meierovich19} Meierovich, B. E. Static State of a Black Hole Supported by Dark Matter.  {\em Universe} {\bf 2019},  5(9), 198.  

\bibitem{Abdikamalov19} Abdikamalov, A. B.; et al.  Black hole mimicker hiding in the shadow: Optical properties of the  $\ensuremath{\gamma}$ metric. {\em Phys. Rev. D} {\bf 2019}, 100,  024014.

\bibitem{Zhu19b} Zhu, T.; et al.  Shadows and deflection angle of charged and slowly rotating black holes in Einstein-\AE{}ther theory. {\em Phys. Rev. D} {\bf 2019}, 100,  044055. 

\bibitem{Tian19} Tian, S. X.; Zhu, Z.-H. Testing the Schwarzschild metric in a strong field region with the Event Horizon Telescope.  {\em Phys. Rev. D} {\bf 2019}, 100,  064011.  

\bibitem{Davoudiasl19} Davoudiasl, H.; Denton, P. B.  Ultralight Boson Dark Matter and Event Horizon Telescope Observations of M87*. {\em Phys. Rev. Lett.} {\bf 2019}, 123, 021102.

\bibitem{Konoplya19b} Konoplya, R. A.; Pappas, T.; Zhidenko, A. Einstein-scalar-Gauss-Bonnet black holes: Analytical approximation for the metric and applications to calculations of shadows. {\em Phys. Rev. D} {\bf 2020}, 101, 044054. 

\bibitem{Hess19} Hess, P. O.; L\'opez-Moreno, E. Kerr Black Holes within a Modified Theory of Gravity. {\em Universe} {\bf 2019}, 5(9), 191. 

\bibitem{Rummel20} Rummel, M.; Burgess, C. P. Constraining Fundamental Physics with the Event Horizon Telescope. {\em arXiv} {\bf 2020}, {arXiv:2001.00041}.

\bibitem{Alexeyev20} Alexeyev, S. O.; Prokopov, V. A. Shadows from spinning black holes in extended gravity. {\em arXiv} {\bf 2020}, {arXiv:2001.09272}.

\bibitem{Chagoya20} Chagoya, J.; et al. Strong lensing by DHOST black holes. {\em arXiv} {\bf 2020}, {arXiv:2007.09473}.

\bibitem{Kardashev14} Kardashev, N. S.; et al. Review of scientific topics for the Millimetron space observatory. {\em Phys. Usp.} {\bf 2014}, 57, 1199–1228;  Usp. Fiz. Nauk {\bf 2014}, 184, 1319-1352. 

\bibitem{Roelofs19} Roelofs, F.; et al.  Simulations of imaging the event horizon of Sagittarius A* from space. {\em Astron. Astrophys.} {\bf 2019}, 625, A124.

\bibitem{Palumbo19} Palumbo, D. C. M.; et al.   Metrics and Motivations for Earth-Space VLBI: Time-resolving Sgr A* with the Event Horizon Telescope.  {\em Astrophys. J} {\bf 2019}, 881, 62.

\bibitem{BPT} Bardeen, J. M.; Press, W. H.; Teukolsky, S. A.  Rotating Black Holes: Locally Nonrotating Frames, Energy Extraction, and Scalar Synchrotron Radiation.  {\em Astrophys. J.} {\bf 1972}, 178, 347-370. 

\bibitem{BoyerLindquist} Boyer, R. H.; Lindquist, R. W. Maximal Analytic Extension of the Kerr Metric.  {\em J. Math. Phys.} {\bf 1967}, 8, 265.

\bibitem{Carter68} Carter, B. Global Structure of the Kerr Family of Gravitational Fields. {\em Phys. Rev.} {\bf 1968}, 174, 1559.

\bibitem{deFelice} de Felice, F. Equatorial geodesic motion in the gravitational field of a rotating source. {\em Nuovo Cimento B} {\bf 1968}, 57, 351–388.

\bibitem{Bardeen70} Bardeen, J. M.  Stability of Circular Orbits in Stationary, Axisymmetric Space-Times. {\em Astrophys. J.} {\bf 1970}, 161,  103-109.

\bibitem{Bardeen70b} Bardeen, J. M. A Variational Principle for Rotating Stars in General Relativity.  {\em Astrophys. J.} {\bf 1970}, 162,  71-95. 

\bibitem{mtw} Misner, C. W.; Thorne, K. S.; Wheeler J. A. {\em Gravitation}; W. H. Freeman: San Francisco, CA, 1973.

\bibitem{Galtsov} Gal'tsov, D. V. {\em Particles and fields in the vicinity of black holes}; Moscow Univ. Press, Moscow, Russia, 1986 (in Russian).

\bibitem{Wilkins} Wilkins, D. C. Bound Geodesics in the Kerr Metric. {\em Phys. Rev. D} {\bf 1972}, 5, 814-822. 

\bibitem{Luminet79} Luminet, J-P. Image of a spherical black hole with thin accretion disk. {\em Astron. Astrophys.} {\bf 1979}, 75, 228-235. 

\bibitem{Zakharov94} Zakharov, A. F. Comment on ``Gravitational lensing of massive particles in Schwarzschild gravity''. {\em Class. Quantum Grav.} {\bf 1994}, 11,  1027. 

\bibitem{Beckwith05} Beckwith, K.; Done, C.  Extreme gravitational lensing near rotating black holes. {\em Mon. Not. R. Astron. Soc.} {\bf 2005}, 359,  1217–1228. 

\bibitem{ZakhPaoIngrNuc05} Zakharov, A. F.; et al. Measuring the black hole parameters in the galactic center with RADIOASTRON.   {\em New Astron.} {\bf 2005}, 10, 479-489.

\bibitem{Takahashi05} Takahashi, R.  Black Hole Shadows of Charged Spinning Black Holes.  {\em Publ. Astron. Soc. Jpn.} {\bf 2005}, 57, 273–277.

\bibitem{Takahashi07} Takahashi, R.; Watarai, K. Eclipsing light curves for accretion flows around a rotating black hole and atmospheric effects of the companion star. {\em Mon. Not. R. Astron. Soc.} {\bf 2007}, 374, 1515–1526. 

\bibitem{Bakala07} Bakala, P.; et al. Extreme gravitational lensing in vicinity of Schwarzschild-de Sitter black holes. {\em Cent. Eur. J. Phys.} {\bf 2007}, 5,  599–610.

\bibitem{Huang07} Huang, L.; et al. Black Hole Shadow Image and Visibility Analysis of Sagittarius A*. {\em Mon. Not. R. Astron. Soc.} {\bf 2007}, 379, 833-840. 

\bibitem{Virbhadra09} Virbhadra, K. S. Relativistic images of Schwarzschild black hole lensing. {\em Phys. Rev. D} {\bf 2009}, 79, 083004.

\bibitem{Hioki09} Hioki, K.; Maeda, K. I.  Measurement of the Kerr Spin Parameter by Observation of a Compact Object's Shadow. {\em Phys. Rev. D} {\bf 2009}, 80,  024042. 

\bibitem{Schee09} Schee, J.; Stuchl\'ik, Z. Profiles of emission lines generated by rings orbiting braneworld Kerr black holes.  {\em Gen. Relativ. Gravit.} {\bf 2009}, 41,  1795–1818.

\bibitem{Dexter09} Dexter, J.;  Agol, E.   A Fast New Public Code for Computing Photon Orbits in a Kerr Spacetime.  {\em Astrophys. J.} {\bf 2009}, 696, 1616-1629. 

\bibitem{JohPsaltis10} Johannsen, T.; Psaltis, D.  Testing the No-Hair Theorem with Observations in the Electromagnetic Spectrum: II. Black-Hole Images.  {\em Astrophys. J.} {\bf 2010}, 718, 446–454. 

\bibitem{Amarilla10} Amarilla, L.;  Eiroa, E. F.; Giribet, G. Null geodesics and shadow of a rotating black hole in extended Chern-Simons modified gravity. {\em Phys. Rev. D} {\bf 2010}, 81, 124045. 

\bibitem{Nitta11} Nitta, D.; Chiba, T.; Sugiyama, N.  Shadows of Colliding Black Holes. {\em Phys. Rev. D} {\bf 2011}, 84, 063008.

\bibitem{Yumoto12} Yumoto, A.; et al. Shadows of Multi-Black Holes: Analytic Exploration. {\em Phys. Rev. D} {\bf 2012}, 86, 103001. 

\bibitem{Abdujabbarov13} Abdujabbarov, A.; et al.  Shadow of Kerr-Taub-NUT black hole. {\em Astrophys. Space Sci.} {\bf 2013}, 344, 429-435. 

\bibitem{Bambi13} Bambi, C. Can the supermassive objects at the centers of galaxies be traversable wormholes? The first test of strong gravity for mm/sub-mm VLBI facilities. {\em Phys. Rev. D} {\bf 2013}, 87, 107501. 

\bibitem{Atamurotov13} Atamurotov, F.; Abdujabbarov, A.; Ahmedov, B. Shadow of rotating Hořava-Lifshitz black hole. {\em Astrophys. Space Sci.} {\bf 2013}, 348, 179–188. 

\bibitem{Atamurotov13b} Atamurotov, F.; Abdujabbarov, A.; Ahmedov, B. Shadow of rotating non-Kerr black hole. {\em Phys. Rev. D} {\bf 2013}, 88,  064004. 

\bibitem{Wei13} Wei, S.-W.; Liu, Y.-X.  Observing the shadow of Einstein-Maxwell-Dilaton-Axion black hole.  {\em JCAP} {\bf 2013}, 11, 063.

\bibitem{Tsukamoto14} Tsukamoto, N.; Li, Z.; Bambi, C. Constraining the spin and the deformation parameters from the black hole shadow.  {\em JCAP} {\bf 2014}, 06, 043. 

\bibitem{Papnoi14} Papnoi, U.; et al. Shadow of five-dimensional rotating Myers-Perry black hole. {\em Phys. Rev. D} {\bf 2014}, 90,  024073.

\bibitem{Tinchev14} Tinchev, V. K.; Yazadjiev, S. S. Possible imprints of cosmic strings in the shadows of galactic black holes.  {\em Int. J. Mod. Phys. D} {\bf 2014}, 23, 1450060. 

\bibitem{Kraniotis14} Kraniotis, G. V.  Gravitational lensing and frame dragging of light in the Kerr–Newman and the Kerr–Newman (anti) de Sitter black hole spacetimes.  {\em Gen. Relativ. Gravit.} {\bf 2014}, 46,  1818. 

\bibitem{Ghas15} Ghasemi-Nodehi, M.; Li, Z.; Bambi, C. Shadows of CPR black holes and tests of the Kerr metric.  {\em Eur. Phys. J. C} {\bf 2015}, 75, 315. 

\bibitem{Tinchev15} Tinchev, V. K. The Shadow of Generalized Kerr Black Holes with Exotic Matter. {\em Chin. J. Phys.} {\bf 2015}, 53,  110113. 

\bibitem{Gralla15} Gralla, S. E.; Porfyriadis, A. P.;  Warburton, N.  Particle on the Innermost Stable Circular Orbit of a Rapidly Spinning Black Hole.  {\em Phys. Rev. D} {\bf 2015}, 92,  064029. 

\bibitem{Atamurotov15} Atamurotov, F.; Ahmedov, B.;  Abdujabbarov, A. Optical properties of black holes in the presence of a plasma: The shadow. {\em Phys. Rev. D} {\bf 2015}, 92, 084005. 

\bibitem{Perlick15} Perlick, V.; Tsupko, O. Y.; Bisnovatyi-Kogan, G. S. Influence of a plasma on the shadow of a spherically symmetric black hole.  {\em Phys. Rev. D} {\bf 2015}, 92,  104031.

\bibitem{Shipley16} Shipley, J.; Dolan, S. R. Binary black hole shadows, chaotic scattering and the Cantor set.  {\em Class. Quantum Grav.} {\bf 2016}, 33, 175001.

\bibitem{Liu16} Liu, X.; Yang, N.; Jia, J. Gravitational lensing of massive particles in Schwarzschild gravity.   {\em Class. Quantum Grav.} {\bf 2016}, 33,  175014.

\bibitem{Yang16} Yang, L.; Li, Z. Shadow of a dressed black hole and determination of spin and viewing angle. {\em Int. J. Mod. Phys. D} {\bf 2016}, 25, 1650026. 

\bibitem{Strom16} Gralla, S. E.; Lupsasca, A.; Strominger, A. Near-horizon Kerr magnetosphere.  {\em Phys. Rev. D} {\bf 2016}, 93,  104041. 

\bibitem{Amir16} Amir, M.; Ghosh, S. G. Shapes of rotating nonsingular black hole shadows. {\em Phys. Rev. D} {\bf 2016}, 94, 024054. 

\bibitem{Gralla16} Gralla, S. E.; Zimmerman. A.; Zimmerman, P. Transient Instability of Rapidly Rotating Black Holes. {\em Phys. Rev. D} {\bf 2016}, 94,  084017. 

\bibitem{Vincent16} Vincent, F. H.; et al. Astrophysical imaging of Kerr black holes with scalar hair. {\em Phys. Rev. D} {\bf 2016}, 94, 084045. 

\bibitem{Dastan16} Dastan, S.; Saffari, R.; Soroushfar, S. Shadow of a Charged Rotating Black Hole in $f(R)$ Gravity. {\em arXiv} {\bf 2016}  {arXiv:1606.06994}.

\bibitem{Tret16} Tretyakova, D. A.; Adyev, T. M. Horndeski/Galileon black hole shadows. {\em arXiv} {\bf 2016}  {arXiv:1610.07300}.

\bibitem{Dastan16b} Dastan, S.; Saffari, R.; Soroushfar, S. Shadow of a Kerr-Sen dilaton-axion Black Hole. {\em arXiv} {\bf 2016}  {arXiv:1610.09477}.

\bibitem{Sharif16} Sharif, M.; Iftikhar, S. Shadow of a charged rotating non-commutative black hole. {\em Eur. Phys. J. C} {\bf 2016}, 76, 630. 

\bibitem{Opatrny17} Opatrn\'y, T.; Richterek, L.; Bakala, P. Life under a black sun. {\em Am. J. Phys.} {\bf 2017}, 85, 14. 

\bibitem{Cunha17b} Cunha, P. V. P.; et al. Shadows of Einstein-dilaton-Gauss-Bonnet black holes. {\em Phys. Lett. B} {\bf 2017}, 768,  373-379. 

\bibitem{Singh17} Singh, B. P.; Ghosh, S. G. Shadow of Schwarzschild-Tangherlini black holes. {\em arXiv} {\bf 2017}, {arXiv:1707.07125}.

\bibitem{Wang17} Wang, M.; Chen, S.; Jing, J.  Shadow casted by a Konoplya-Zhidenko rotating non-Kerr black hole. {\em JCAP} {\bf 2017}, 10, 051.

\bibitem{Amir17} Amir, M.; Singh, B. P.; Ghosh, S. G. Shadows of rotating five-dimensional charged EMCS black holes.  {\em Eur. Phys. J. C} {\bf 2018}, 78, 399. 

\bibitem{Strom17} Porfyriadis, A. P.; Shi, Y.; Strominger A. Photon emission near extreme Kerr black holes. {\em Phys. Rev. D} {\bf 2017}, 95, 064009. 

\bibitem{Strom18} Gralla, S. E.; Lupsasca, A.; Strominger, A. Observational Signature of High Spin at the Event Horizon Telescope.  {\em Mon. Not. R. Astron. Soc.}  {\bf 2017}, 475,  3829-3853. 

\bibitem{Tsupko17a} Perlick, V.; Tsupko, O. Y. Light propagation in a plasma on Kerr spacetime: Separation of the Hamilton-Jacobi equation and calculation of the shadow.  {\em Phys. Rev. D} {\bf 2017}, 95,  104003. 

\bibitem{Tsupko17b} Tsupko, O. Y. Analytical calculation of black hole spin using deformation of the shadow.  {\em Phys. Rev. D} {\bf 2017}, 95, 104058. 

\bibitem{BisnovatyiTsupko17} Bisnovatyi-Kogan, G. S.; Tsupko, O. Y. Gravitational Lensing in Presence of Plasma: Strong Lens Systems, Black Hole Lensing and Shadow. {\em Universe} {\bf 2017}, 3(3), 57.

\bibitem{Stuchlik18} Stuchl\'ik, Z.; Charbul\'ak, D. Light escape cones in local reference frames of Kerr–de Sitter  black hole spacetimes and related black hole shadows. {\em Eur. Phys. J. C} {\bf 2018}, 78,  180. 

\bibitem{Cunha18b} Cunha, P. V. P.; Herdeiro, C. A. R.; Rodriguez, M. J. Does the black hole shadow probe the event horizon geometry? {\em Phys. Rev. D} {\bf 2018}, 97,  084020. 

\bibitem{Huang18} Huang, Y.; Dong, Y.-P.; Liu, D.-J. Revisiting the shadow of a black hole in the presence of a plasma. {\em Int. J. Mod. Phys. D} {\bf 2018}, 27, 1850114. 

\bibitem{Tsukamoto18} Tsukamoto, N. Black hole shadow in an asymptotically flat, stationary, and axisymmetric spacetime: The Kerr-Newman and rotating regular black holes. {\em Phys. Rev. D} {\bf 2018}, 97, 064021. 

\bibitem{Bisnovatyi18} Bisnovatyi-Kogan, G. S.; Tsupko, O. Y. Shadow of a black hole at cosmological distances. {\em Phys. Rev. D} {\bf 2018}, 98, 084020. 

\bibitem{Hou18} Hou, X.; Xu, Z.; Wang, J. Rotating Black Hole Shadow in Perfect Fluid Dark Matter.  {\em JCAP} {\bf 2018}, 12, 040. 

\bibitem{Yan19} Yan, H. Influence of a plasma on the observational signature of a high-spin Kerr black hole.  {\em Phys. Rev. D} {\bf 2019}, 99, 084050. 

\bibitem{Vagnozzi19} Vagnozzi, S.; Visinelli.; L. Hunting for extra dimensions in the shadow of M87*. {\em Phys. Rev. D} {\bf 2019}, 100,  024020. 

\bibitem{Gyulchev19} Gyulchev, G.; et al. Image of the Janis-Newman-Winicour naked singularity with a thin accretion disk. {\em Phys. Rev. D} {\bf 2019}, 100,  024055. 

\bibitem{Kumar19} Kumar, R.; Ghosh, S. G.; Wang, A. Shadow cast and deflection of light by charged rotating regular black holes. {\em Phys. Rev. D} {\bf 2019}, 100, 124024. 

\bibitem{Konoplya19} Konoplya, R. A. Shadow of a black hole surrounded by dark matter.  {\em Phys. Lett. B} {\bf 2019}, 795, 1-6. 

\bibitem{Sabir19} Ali, M. S.; Amir, M. Shadow of rotating charged black hole with Weyl corrections. {\em arXiv} {\bf 2019}, {arXiv:1906.04146}.

\bibitem{Johnson19} Johnson, M. D.; et al. Universal interferometric signatures of a black hole’s photon ring. {\em Sci. Adv.} {\bf 2020}, 6(12), eaaz1310.

\bibitem{Siino19} Siino, M. Generalization of Photon Sphere by Causal Structure--- In Order to See Dynamical Black Hole Shadow. {\em arXiv} {\bf 2019}, {arXiv:1908.02921}.

\bibitem{Zhang19} Zhang, M.; Guo, M. Can shadows reflect phase structures of black holes? {\em arXiv} {\bf 2019}, {arXiv:1909.07033}.

\bibitem{Shipley19} Shipley, J. O. Strong-field gravitational lensing by black holes. {\em arXiv} {\bf 2019}, {arXiv:1909.04691}.   

\bibitem{Shaikh19} Shaikh, R.; et al. Shadows of spherically symmetric black holes and naked singularities.  {\em Mon. Not. R. Astron. Soc.} {\bf 2019}, 482, 52-64. 

\bibitem{Shaikh19b} Shaikh, R.; Joshi, P. S. Can we distinguish black holes from naked singularities by the images of their accretion disks? {\em JCAP} {\bf 2019}, 10, 064. 

\bibitem{Ding19} Ding, C.; et al. Exact Kerr-like solution and its shadow in a gravity model with spontaneous Lorentz symmetry breaking. {\em Eur. Phys. J. C } {\bf 2020}, 80, 178.

\bibitem{Narayan19} Narayan, R.; Johnson, M. D.; Gammie, C. F. The Shadow of a Spherically Accreting Black Hole.  {\em Astrophys. J. Lett.} {\bf 2019}, 885, L33. 

\bibitem{Goddi19} Goddi, C.; et al. First M87 Event Horizon Telescope Results and the Role of ALMA.  {\em The Messenger} {\bf 2019}, 177, 25. 

\bibitem{Feng19} Feng, X.-H.; Lu, H. On the Size of Rotating Black Holes. {\em The European Physical Journal C} {\bf 2020}, 80, 551 

\bibitem{Allahyari19} Allahyari, A.; et al. Magnetically charged black holes from non-linear electrodynamics and the Event Horizon Telescope. {\em JCAP} {\bf 2020}, 02, 003.

\bibitem{Konoplya19c} Konoplya, R. A. Quantum corrected black holes: Quasinormal modes, scattering, shadows.  {\em Phys. Lett. B} {\bf 2020}, 804, 135363. 

\bibitem{doknazsm19} Dokuchaev, V. I.; Nazarova, N. O.; Smirnov, V. P. Event horizon silhouette: implications to supermassive black holes in the galaxies M87 and Milky Way. {\em Gen. Relativ. Gravit.} {\bf 2019}, 51,  81. 

\bibitem{dokuch19} Dokuchaev, V. I. To see the invisible: Image of the event horizon within the black hole shadow.  {\em Int. J. Mod. Phys. D} {\bf 2019}, 28, 1941005.

\bibitem{Cunha20} Cunha, P. V. P.; et al. Lensing and shadow of a black hole surrounded by a heavy accretion disk. {\em JCAP} {\bf 2020}, 03,  035.

\bibitem{Tsupko20} Tsupko, O. Y.; Bisnovatyi-Kogan, G. S. Hills and holes in the microlensing light curve due to plasma environment around gravitational lens.   {\em Mon. Not. R. Astron. Soc.} {\bf 2020}, 491, 5636–5649. 

\bibitem{Vagnozzi20} Vagnozzi, S.; Bambi, C.; Visinelli, L.  Concerns regarding the use of black hole shadows as standard rulers. {\em Class. Quanum Grav.} {\bf 2020}, 37, 087001. 

\bibitem{Yu20} Yu, S.; Gao, C. An exact black hole spacetime with scalar field and its shadow together with quasinormal modes. {\em arXiv} {\bf 2020}, {arXiv:2001.01137}.

\bibitem{Banerjee20} Banerjee, I.; Chakraborty, S.; SenGupta, S. Silhouette of M87*: a new window to peek into the world of hidden dimensions.  {\em Phys. Rev. D}  {\bf 2020}, 101, 041301. 

\bibitem{gralla20} Gralla, S.; Lupsasca, A. Lensing by Kerr black holes.  {\em Phys. Rev. D}  {\bf 2020}, 101, 044031. 

\bibitem{Chang20} Chang, Z.; Zhu, Q.-H.  Revisiting a rotating black hole shadow with astrometric observables. {\em Phys. Rev. D} {\bf 2020}, 101,  084029. 

\bibitem{Himwich20} Himwich, E.; et al. Universal polarimetric signatures of the black hole photon ring.  {\em Phys. Rev. D} {\bf 2020}, 101,  084020. 

\bibitem{Li20} Li, P.-C.; Guo, M.; Chen, B. Shadow of a spinning black hole in an expanding universe. {\em Phys. Rev. D} {\bf 2020}, 101,  084041. 

\bibitem{Jusufi20} Jusufi, K. Quasinormal modes of black holes surrounded by dark matter and their connection with the shadow radius.  {\em Phys. Rev. D} {\bf 2020}, 101, 084085.

\bibitem{Bakala20} Bakala, P.;  Do\v cekal, J.; Turo\v nova, Z. Habitable zones around almost extremely spinning black holes (black sun revisited). {\em Astrophys. J.} {\bf 2020}, 889,  41.

\bibitem{Anantua20} Anantua, R.; Ressler, S.; Quataert,  E. On the Comparison of AGN with GRMHD Simulations: I. Sgr A*. {\em Mon. Not. R. Astron. Soc.} {\bf 2020}, 493,  1404-1418.

\bibitem{Belhaj20} Belhaj, A. et al. Shadows of Charged and Rotating Black Holes with a Cosmological Constant. {\em arXiv} {\bf 2020}, {arXiv:2007.09058}.

\bibitem{Grenzebach14} Grenzebach, A.; Perlick, V.;  L\"ammerzahl, C. Photon regions and shadows of Kerr-Newman-NUT black holes with a cosmological constant. {\em Phys. Rev. D} {\bf 2014}, 89, 124004. 

\bibitem{Grenzebach15} Grenzebach, A.; Perlick, V.;  L\"ammerzahl, C. Photon regions and shadows of accelerated black holes.  {\em Int. J. Mod. Phys. D}  {\bf 2015}, 24, 1542024. 

\bibitem{Cunha18a} Cunha, P. V. P.; Herdeiro, C. A. R. Shadows and strong gravitational lensing: a brief review. {\em Gen. Relativ. Gravit.} {\bf 2018}, 50, 42-69. 

\bibitem{CunnBardeen72} Cunningham, C. T.; Bardeen, J. M. The Optical Appearance of a Star Orbiting an Extreme Kerr Black Hole.   {\em Astrophys. J.} {\bf 1972}, 173, L137. 

\bibitem{CunnBardeen73} Cunningham, C. T.; Bardeen, J. M. The Optical Appearance of a Star Orbiting an Extreme Kerr Black Hole. {\em Astrophys. J.} {\bf 1973}, 183, 237-264. 

\bibitem{Viergutz93} Viergutz, S. U. Image generation in Kerr geometry. I. Analytical investigations on the stationary emitter-observer problem. {\em Astron. Astrophys.} {\bf 1993}, 272, 355-377. 

\bibitem{RauchBlandf94} Rauch, K. P.; Blandford, R. D. Optical Caustics in a Kerr Spacetime and the Origin of Rapid X-Ray Variability in Active Galactic Nuclei. {\em Astrophys J.} {\bf 1994}, 421, 46-68. 

\bibitem{GralHolzWald19} Gralla, S. E.; Holz, D. E.; Wald, R. M. Black hole shadows, photon rings, and lensing rings.  {\em Phys. Rev. D} {\bf 2019}, 100,  024018. 

\bibitem{doknaz17} Dokuchaev, V. I.; Nazarova, N. O. Gravitational lensing of a star by a rotating black hole. {\em JETP Lett.} {\bf 2017}, 106,  637–-642;  {\em Pis'ma Zh. Eksp. Teor. Fiz.} {\bf 2017}, 106,  609–-614. 

\bibitem{doknaz18b} Dokuchaev, V. I.; Nazarova, N. O. Star motion around rotating black hole.  {https://youtu.be/P6DneV0vk7U}. 

\bibitem{Claudel01} Claudel, C-M.; Virbhadra, K. S.; Ellis, G. F. 
The geometry of photon surfaces.  {\em J. Math. Phys.} {\bf 2001}, 42, 818--838. 

\bibitem{Grossman12} Grossman, R.; Levin, J.; Perez-Giz, G. 
The harmonic structure of generic Kerr orbits.  {\em Phys. Rev. D} {\bf 2012}, 85, 023012. 

\bibitem{Hod13} Hod, S. Spherical null geodesics of rotating Kerr black holes.  {\em Phys. Lett. B} {\bf 2013}, 718,  1552-1556. 

\bibitem{Liu19} Liu, C.; Ding, C.; Jing, J. Selected spherical photon orbits around a deformed Kerr black hole. {\em Sci. China Phys. Mech. Astron.} {\bf 2019}, 62, 010411. 

\bibitem{Glampedakis19} Glampedakis, K.; Pappas, G. Modification of photon trapping orbits as a diagnostic of non-Kerr spacetimes. {\em Phys. Rev. D} {\bf 2019}, 99, 124041.

\bibitem{Hughes19} Hughes, S. A Nearly horizon skimming orbits of Kerr black holes. {\em Phys. Rev. D} {\bf 2019}, 63, 064016.

\bibitem{Teo20} Teo, E. Spherical orbits around a Kerr black hole. {\em arXiv} {\bf 2020}, arXiv:2007.04022.

\bibitem{Psaltis15} Psaltis, D.; et al. Event-Horizon-Telescope Evidence for Alignment of the Black Hole in the Center of the Milky Way with the Inner Stellar Disk. {\em Astrophys. J.} {\bf 2015}, 798, 15. 

\bibitem{doknaz20} Dokuchaev, V. I.; Nazarova, N. O. Silhouettes of invisible black holes. {\em Phys. Usp. } {\bf 2020}, 63. (6). 

\bibitem{Lin20} Lin, J. Y-Yu.; et al. Feature Extraction on Synthetic Black Hole Images. {\em arXiv} {\bf 2020}, arXiv:2007.00794.

\bibitem{Bromley97} Bromley, B. C.; Chen K.; Miller, W. A. Line emission from an accretion disk around a rotating black hole: toward a measurement of frame dragging {\em Astrophys. J.} {\bf 1997},475, 57--64.

\bibitem{Fanton97} Fanton, C.; et al. Detecting accretion disks in Active Galactic Nuclei. {\em Publ. Astron. Soc. Jpn.} {\bf 1997}, 49, 159--169.

\bibitem{Fukue03} Fukue, J.  Silhouette of a dressed black hole. {\em Publ. Astron. Soc. Jpn.} {\bf 2003}, 55, 155--159.

\bibitem{Fukue03b} Fukue J. Light-curve diagnosis of a hot spot for accretion-disk models {\em Publ. Astron. Soc. Jpn.} {\bf 2003}, 55, 1121--1125.

\bibitem{Dexter09b} Dexter, J.;, Agol, E.; Fragile, P. C. Millimeter flares and VLBI visibilities from relativistic simulations of magnetized
accretion onto the Galactic Center black hole {\em Astrophys. J.} {\bf 2009}, 703, L142-L146.

\bibitem{Ru-SenLu16} Lu, R-S.; et al. Imaging an event horizon: mitigation of source variability of Sagittarius A*. {\em Astrophys. J.} {\bf 2016}, 817, 173 (8pp).

\bibitem{Luminet19} Luminet, J-P. An illustrated history of black hole imaging : personal recollections (1972--2002). {\em arXiv} {\bf 2019}, arXiv:1902.11196.

\bibitem{Shiokawa19} Shiokawa, H. {\em 2019},  https://eventhorizontelescope.org/simulations-gallery.

\bibitem{doknaz19} Dokuchaev, V. I.; Nazarova, N. O. Event horizon image within black hole shadow. {\em JETP} {\bf 2019}, 128, 578--585. 

\bibitem{Gucht19} van der Gucht, J.; et al. Deep horizon: A machine learning network that recovers accreting black hole parameters. {\em Astron. Astrophys.} {\bf 2020}, 636, A94, 12pp.

\bibitem{Kawashima19}Kawashima, T.; Kino, M.; Akiyama, K. Black Hole Spin Signature in the Black Hole Shadow of M87 in the Flaring State. {\em Astrophys. J.} {\bf 2019}, 878, 27 (9pp).

\bibitem{White20} White, C. J.; et al. The Effects of tilt on the images of black hole accretion flows. {\em Astrophys. J.} {\bf 2020}, 894, 14. 

\bibitem{doknaz18c} Dokuchaev, V. I.; Nazarova, N. O. Infall of the star into rotating black hole viewed by a distant observer.  {https://youtu.be/fps-3frL0AM}. 

\bibitem{Walker18} Walker, R. C.; et al. The Structure and Dynamics of the Sub-parsec Scale Jet in M87 Based on 50 VLBA Observations Over 17 Years at 43 GHz. {\em Astrophys. J.} {\bf 2018}, 855, 128. 

\bibitem{Nalewajko20} Nalewajko, K.; Sikora, M.,; R\'oz\`ansk\`a, A. On the orientation of the crescent image of M87*. {\em Astron. Astrophys.} 
{\bf 2020}, 634, A38. 

\bibitem{doknaz19b} Dokuchaev, V. I.; Nazarova, N. O. The Brightest Point in Accretion Disk and Black Hole Spin: Implication to the Image of Black Hole M87*. {\em Universe} {\bf 2019}, 5(8), 183. 
	
\end{thebibliography}
\end{document}